\let\arxiv\relax
\ifx\arxiv\relax
\let\extended\relax
\fi
\ifx\extended\relax
\documentclass[11pt,a4size]{article}
\pdfoutput=1 
\usepackage{authblk}
\usepackage[numbers]{natbib}
\usepackage{epigraph}
\usepackage{hyperref}
\newcommand\blfootnote[1]{%
  \begingroup
  \renewcommand\thefootnote{}\footnote{#1}%
  \addtocounter{footnote}{-1}%
  \endgroup
}
\else
\documentclass[sigconf]{acmart}
\fi

\pdfoutput=1

\usepackage{amsmath,amssymb,amsfonts}
\usepackage{graphicx}
\usepackage{textcomp}
\usepackage{xcolor}
\usepackage{algorithm}
\usepackage{algpseudocode}
\usepackage{listings}
\usepackage{color}
\usepackage{seqsplit}
\usepackage{endnotes}
\usepackage{needspace}
\usepackage{rotating}
\usepackage{pdflscape}
\usepackage{afterpage}

\definecolor{dkgreen}{rgb}{0,0.6,0}
\definecolor{gray}{rgb}{0.5,0.5,0.5}
\definecolor{mauve}{rgb}{0.58,0,0.82}

\usepackage{url}

\definecolor{background}{HTML}{EEEEEE}

\ifx\arxiv\undefined
\graphicspath{{.}{../diagrams/}}
\fi

\AtBeginDocument{%
  \providecommand\BibTeX{{%
    \normalfont B\kern-0.5em{\scshape i\kern-0.25em b}\kern-0.8em\TeX}}}

\ifx\extended\undefined
\copyrightyear{2020}
\acmYear{2020}
\setcopyright{acmcopyright}
\acmConference[MSR '20]{17th International Conference on Mining Software Repositories}{October 5--6, 2020}{Seoul, Republic of Korea}
\acmBooktitle{17th International Conference on Mining Software Repositories (MSR '20), October 5--6, 2020, Seoul, Republic of Korea}
\acmPrice{15.00}
\acmDOI{10.1145/3379597.3387496}
\acmISBN{978-1-4503-7517-7/20/05}
\fi

\begin{document}

\ifx\extended\relax

\author[1]{Diomidis Spinellis}
\author[1]{Zoe Kotti}
\author[2]{Audris Mockus}
\renewcommand\Affilfont{\itshape\small}
\affil[1]{Athens University of Economics and Business}
\affil[2]{University of Tennessee}
\title{A Dataset for GitHub Repository Deduplication:\\
Extended Description}

\else

\title{A Dataset for GitHub Repository Deduplication}

\author{Diomidis Spinellis}
\email{{dds,zoekotti}@aueb.gr}
\orcid{0000-0003-4231-1897}
\author{Zoe Kotti}
\orcid{0000-0003-3816-9162}
\affiliation{%
 \institution{Athens University of Economics and Business}
}
\author{Audris Mockus}
\email{audris@mockus.org}
\affiliation{%
\institution{University of Tennessee}
}

\renewcommand{\shortauthors}{Spinellis, Kotti, and Mockus}
\fi

\makeatletter
\renewcommand*\makeenmark{\hbox{\textsuperscript{G\theenmark}}}

\newcommand{\kwlist}{Deduplication, fork, project clone, GitHub, dataset}

\newwrite\diagrams
\immediate\openout\diagrams=diagrams.txt

\newcounter{sgraph}
\newcommand{\graphnum}[1]{%
  \ifcsname graphnum@#1\endcsname
  \else
    \stepcounter{sgraph}%
    \expandafter\xdef\csname graphnum@#1\endcsname{\thesgraph}%
    \write\diagrams{\csname graphnum@#1\endcsname\space #1}%
  \fi%
  \csname graphnum@#1\endcsname}

\ifx\extended\relax
\newcommand{\graphref}[1]{%
\footnote{Figure~\ref{fig:#1} and G\graphnum{#1}.}%
}
\else
\newcommand{\graphref}[1]{%
\textsuperscript{G\graphnum{#1}}%
}
\fi

\ifx\extended\relax
\newcommand{\tblSharedCommits}{\emph{shared commits} (184\,526\,310 records)}
\newcommand{\ptblrecSharedCommits}{(184\,526\,310 records)}
\newcommand{\ptblSharedCommits}{(Table \emph{shared commits}---184\,526\,310 records---Listing~\ref{l:SharedCommits})}
\newcommand{\tblBlacklistedProjects}{\emph{blacklisted projects} (2\,341\,896 records)}
\newcommand{\ptblrecBlacklistedProjects}{(2\,341\,896 records)}
\newcommand{\ptblBlacklistedProjects}{(Table \emph{blacklisted projects}---2\,341\,896 records---Listing~\ref{l:BlacklistedProjects})}
\newcommand{\tblMegaCluster}{\emph{mega cluster} (1 records)}
\newcommand{\ptblrecMegaCluster}{(1 records)}
\newcommand{\ptblMegaCluster}{(Table \emph{mega cluster}---1 records---Listing~\ref{l:MegaCluster})}
\newcommand{\tblMostStarredInGroup}{\emph{most starred in group} (7\,553\,705 records)}
\newcommand{\ptblrecMostStarredInGroup}{(7\,553\,705 records)}
\newcommand{\ptblMostStarredInGroup}{(Table \emph{most starred in group}---7\,553\,705 records---Listing~\ref{l:MostStarredInGroup})}
\newcommand{\tblHighestMeanInGroup}{\emph{highest mean in group} (7\,553\,705 records)}
\newcommand{\ptblrecHighestMeanInGroup}{(7\,553\,705 records)}
\newcommand{\ptblHighestMeanInGroup}{(Table \emph{highest mean in group}---7\,553\,705 records---Listing~\ref{l:HighestMeanInGroup})}
\newcommand{\tblFilteredProjectsSharingCommits}{\emph{filtered projects sharing commits} (43\,843\,142 records)}
\newcommand{\ptblrecFilteredProjectsSharingCommits}{(43\,843\,142 records)}
\newcommand{\ptblFilteredProjectsSharingCommits}{(Table \emph{filtered projects sharing commits}---43\,843\,142 records---Listing~\ref{l:FilteredProjectsSharingCommits})}
\newcommand{\tblAllProjectMetrics}{\emph{all project metrics} (125\,486\,232 records)}
\newcommand{\ptblrecAllProjectMetrics}{(125\,486\,232 records)}
\newcommand{\ptblAllProjectMetrics}{(Table \emph{all project metrics}---125\,486\,232 records---Listing~\ref{l:AllProjectMetrics})}
\newcommand{\tblRemovedProjects}{\emph{removed projects} (39\,675\,015 records)}
\newcommand{\ptblrecRemovedProjects}{(39\,675\,015 records)}
\newcommand{\ptblRemovedProjects}{(Table \emph{removed projects}---39\,675\,015 records---Listing~\ref{l:RemovedProjects})}
\newcommand{\tblProjectNcommits}{\emph{project ncommits} (100\,366\,312 records)}
\newcommand{\ptblrecProjectNcommits}{(100\,366\,312 records)}
\newcommand{\ptblProjectNcommits}{(Table \emph{project ncommits}---100\,366\,312 records---Listing~\ref{l:ProjectNcommits})}
\newcommand{\tblDeduplicateByMean}{\emph{deduplicate by mean} (10\,649\,348 records)}
\newcommand{\ptblrecDeduplicateByMean}{(10\,649\,348 records)}
\newcommand{\ptblDeduplicateByMean}{(Table \emph{deduplicate by mean}---10\,649\,348 records---Listing~\ref{l:DeduplicateByMean})}
\newcommand{\tblProjectsSharingCommits}{\emph{projects sharing commits} (44\,380\,204 records)}
\newcommand{\ptblrecProjectsSharingCommits}{(44\,380\,204 records)}
\newcommand{\ptblProjectsSharingCommits}{(Table \emph{projects sharing commits}---44\,380\,204 records---Listing~\ref{l:ProjectsSharingCommits})}
\newcommand{\tblProjectPullRequests}{\emph{project pull requests} (7\,143\,570 records)}
\newcommand{\ptblrecProjectPullRequests}{(7\,143\,570 records)}
\newcommand{\ptblProjectPullRequests}{(Table \emph{project pull requests}---7\,143\,570 records---Listing~\ref{l:ProjectPullRequests})}
\newcommand{\tblProjectGroupSize}{\emph{project group size} (18\,203\,053 records)}
\newcommand{\ptblrecProjectGroupSize}{(18\,203\,053 records)}
\newcommand{\ptblProjectGroupSize}{(Table \emph{project group size}---18\,203\,053 records---Listing~\ref{l:ProjectGroupSize})}
\newcommand{\tblAllProjectMeanMetric}{\emph{all project mean metric} (125\,486\,232 records)}
\newcommand{\ptblrecAllProjectMeanMetric}{(125\,486\,232 records)}
\newcommand{\ptblAllProjectMeanMetric}{(Table \emph{all project mean metric}---125\,486\,232 records---Listing~\ref{l:AllProjectMeanMetric})}
\newcommand{\tblProjectStars}{\emph{project stars} (10\,317\,662 records)}
\newcommand{\ptblrecProjectStars}{(10\,317\,662 records)}
\newcommand{\ptblProjectStars}{(Table \emph{project stars}---10\,317\,662 records---Listing~\ref{l:ProjectStars})}
\newcommand{\tblGroupSize}{\emph{group size} (2\,472\,758 records)}
\newcommand{\ptblrecGroupSize}{(2\,472\,758 records)}
\newcommand{\ptblGroupSize}{(Table \emph{group size}---2\,472\,758 records---Listing~\ref{l:GroupSize})}
\newcommand{\tblMostRecentCommit}{\emph{most recent commit} (100\,366\,312 records)}
\newcommand{\ptblrecMostRecentCommit}{(100\,366\,312 records)}
\newcommand{\ptblMostRecentCommit}{(Table \emph{most recent commit}---100\,366\,312 records---Listing~\ref{l:MostRecentCommit})}
\newcommand{\tblProjectMeanMetric}{\emph{project mean metric} (18\,203\,053 records)}
\newcommand{\ptblrecProjectMeanMetric}{(18\,203\,053 records)}
\newcommand{\ptblProjectMeanMetric}{(Table \emph{project mean metric}---18\,203\,053 records---Listing~\ref{l:ProjectMeanMetric})}
\newcommand{\tblMostForkedInGroup}{\emph{most forked in group} (7\,553\,705 records)}
\newcommand{\ptblrecMostForkedInGroup}{(7\,553\,705 records)}
\newcommand{\ptblMostForkedInGroup}{(Table \emph{most forked in group}---7\,553\,705 records---Listing~\ref{l:MostForkedInGroup})}
\newcommand{\tblNoTable}{\emph{no table} (0 records)}
\newcommand{\ptblrecNoTable}{(0 records)}
\newcommand{\ptblNoTable}{(Table \emph{no table}---0 records---Listing~\ref{l:NoTable})}
\newcommand{\tblDeduplicateByStars}{\emph{deduplicate by stars} (10\,649\,348 records)}
\newcommand{\ptblrecDeduplicateByStars}{(10\,649\,348 records)}
\newcommand{\ptblDeduplicateByStars}{(Table \emph{deduplicate by stars}---10\,649\,348 records---Listing~\ref{l:DeduplicateByStars})}
\newcommand{\tblNoiseProjects}{\emph{noise projects} (37\,333\,119 records)}
\newcommand{\ptblrecNoiseProjects}{(37\,333\,119 records)}
\newcommand{\ptblNoiseProjects}{(Table \emph{noise projects}---37\,333\,119 records---Listing~\ref{l:NoiseProjects})}
\newcommand{\tblMeanStarMismatch}{\emph{mean star mismatch} (534\,903 records)}
\newcommand{\ptblrecMeanStarMismatch}{(534\,903 records)}
\newcommand{\ptblMeanStarMismatch}{(Table \emph{mean star mismatch}---534\,903 records---Listing~\ref{l:MeanStarMismatch})}
\newcommand{\tblProjectMetrics}{\emph{project metrics} (18\,203\,053 records)}
\newcommand{\ptblrecProjectMetrics}{(18\,203\,053 records)}
\newcommand{\ptblProjectMetrics}{(Table \emph{project metrics}---18\,203\,053 records---Listing~\ref{l:ProjectMetrics})}
\newcommand{\tblProjectForks}{\emph{project forks} (6\,958\,551 records)}
\newcommand{\ptblrecProjectForks}{(6\,958\,551 records)}
\newcommand{\ptblProjectForks}{(Table \emph{project forks}---6\,958\,551 records---Listing~\ref{l:ProjectForks})}
\newcommand{\tblAcgroups}{\emph{acgroups} (18\,203\,053 records)}
\newcommand{\ptblrecAcgroups}{(18\,203\,053 records)}
\newcommand{\ptblAcgroups}{(Table \emph{acgroups}---18\,203\,053 records---Listing~\ref{l:Acgroups})}
\newcommand{\tblForksClonesNoise}{\emph{forks clones noise} (50\,324\,363 records)}
\newcommand{\ptblrecForksClonesNoise}{(50\,324\,363 records)}
\newcommand{\ptblForksClonesNoise}{(Table \emph{forks clones noise}---50\,324\,363 records---Listing~\ref{l:ForksClonesNoise})}
\newcommand{\tblProjectIssues}{\emph{project issues} (9\,498\,704 records)}
\newcommand{\ptblrecProjectIssues}{(9\,498\,704 records)}
\newcommand{\ptblProjectIssues}{(Table \emph{project issues}---9\,498\,704 records---Listing~\ref{l:ProjectIssues})}
\newcommand{\tblCommitPercentiles}{\emph{commit percentiles} (32 records)}
\newcommand{\ptblrecCommitPercentiles}{(32 records)}
\newcommand{\ptblCommitPercentiles}{(Table \emph{commit percentiles}---32 records---Listing~\ref{l:CommitPercentiles})}
\newcommand{\tblDeduplicateByForks}{\emph{deduplicate by forks} (10\,649\,348 records)}
\newcommand{\ptblrecDeduplicateByForks}{(10\,649\,348 records)}
\newcommand{\ptblDeduplicateByForks}{(Table \emph{deduplicate by forks}---10\,649\,348 records---Listing~\ref{l:DeduplicateByForks})}

\else
\newcommand{\tblSharedCommits}{\emph{shared commits} (184\,526\,310 records)}
\newcommand{\ptblrecSharedCommits}{(184\,526\,310 records)}
\newcommand{\ptblSharedCommits}{(Table \emph{shared commits}---184\,526\,310 records)}
\newcommand{\tblBlacklistedProjects}{\emph{blacklisted projects} (2\,341\,896 records)}
\newcommand{\ptblrecBlacklistedProjects}{(2\,341\,896 records)}
\newcommand{\ptblBlacklistedProjects}{(Table \emph{blacklisted projects}---2\,341\,896 records)}
\newcommand{\tblMegaCluster}{\emph{mega cluster} (1 records)}
\newcommand{\ptblrecMegaCluster}{(1 records)}
\newcommand{\ptblMegaCluster}{(Table \emph{mega cluster}---1 records)}
\newcommand{\tblMostStarredInGroup}{\emph{most starred in group} (7\,553\,705 records)}
\newcommand{\ptblrecMostStarredInGroup}{(7\,553\,705 records)}
\newcommand{\ptblMostStarredInGroup}{(Table \emph{most starred in group}---7\,553\,705 records)}
\newcommand{\tblHighestMeanInGroup}{\emph{highest mean in group} (7\,553\,705 records)}
\newcommand{\ptblrecHighestMeanInGroup}{(7\,553\,705 records)}
\newcommand{\ptblHighestMeanInGroup}{(Table \emph{highest mean in group}---7\,553\,705 records)}
\newcommand{\tblFilteredProjectsSharingCommits}{\emph{filtered projects sharing commits} (43\,843\,142 records)}
\newcommand{\ptblrecFilteredProjectsSharingCommits}{(43\,843\,142 records)}
\newcommand{\ptblFilteredProjectsSharingCommits}{(Table \emph{filtered projects sharing commits}---43\,843\,142 records)}
\newcommand{\tblAllProjectMetrics}{\emph{all project metrics} (125\,486\,232 records)}
\newcommand{\ptblrecAllProjectMetrics}{(125\,486\,232 records)}
\newcommand{\ptblAllProjectMetrics}{(Table \emph{all project metrics}---125\,486\,232 records)}
\newcommand{\tblRemovedProjects}{\emph{removed projects} (39\,675\,015 records)}
\newcommand{\ptblrecRemovedProjects}{(39\,675\,015 records)}
\newcommand{\ptblRemovedProjects}{(Table \emph{removed projects}---39\,675\,015 records)}
\newcommand{\tblProjectNcommits}{\emph{project ncommits} (100\,366\,312 records)}
\newcommand{\ptblrecProjectNcommits}{(100\,366\,312 records)}
\newcommand{\ptblProjectNcommits}{(Table \emph{project ncommits}---100\,366\,312 records)}
\newcommand{\tblDeduplicateByMean}{\emph{deduplicate by mean} (10\,649\,348 records)}
\newcommand{\ptblrecDeduplicateByMean}{(10\,649\,348 records)}
\newcommand{\ptblDeduplicateByMean}{(Table \emph{deduplicate by mean}---10\,649\,348 records)}
\newcommand{\tblProjectsSharingCommits}{\emph{projects sharing commits} (44\,380\,204 records)}
\newcommand{\ptblrecProjectsSharingCommits}{(44\,380\,204 records)}
\newcommand{\ptblProjectsSharingCommits}{(Table \emph{projects sharing commits}---44\,380\,204 records)}
\newcommand{\tblProjectPullRequests}{\emph{project pull requests} (7\,143\,570 records)}
\newcommand{\ptblrecProjectPullRequests}{(7\,143\,570 records)}
\newcommand{\ptblProjectPullRequests}{(Table \emph{project pull requests}---7\,143\,570 records)}
\newcommand{\tblProjectGroupSize}{\emph{project group size} (18\,203\,053 records)}
\newcommand{\ptblrecProjectGroupSize}{(18\,203\,053 records)}
\newcommand{\ptblProjectGroupSize}{(Table \emph{project group size}---18\,203\,053 records)}
\newcommand{\tblAllProjectMeanMetric}{\emph{all project mean metric} (125\,486\,232 records)}
\newcommand{\ptblrecAllProjectMeanMetric}{(125\,486\,232 records)}
\newcommand{\ptblAllProjectMeanMetric}{(Table \emph{all project mean metric}---125\,486\,232 records)}
\newcommand{\tblProjectStars}{\emph{project stars} (10\,317\,662 records)}
\newcommand{\ptblrecProjectStars}{(10\,317\,662 records)}
\newcommand{\ptblProjectStars}{(Table \emph{project stars}---10\,317\,662 records)}
\newcommand{\tblGroupSize}{\emph{group size} (2\,472\,758 records)}
\newcommand{\ptblrecGroupSize}{(2\,472\,758 records)}
\newcommand{\ptblGroupSize}{(Table \emph{group size}---2\,472\,758 records)}
\newcommand{\tblMostRecentCommit}{\emph{most recent commit} (100\,366\,312 records)}
\newcommand{\ptblrecMostRecentCommit}{(100\,366\,312 records)}
\newcommand{\ptblMostRecentCommit}{(Table \emph{most recent commit}---100\,366\,312 records)}
\newcommand{\tblProjectMeanMetric}{\emph{project mean metric} (18\,203\,053 records)}
\newcommand{\ptblrecProjectMeanMetric}{(18\,203\,053 records)}
\newcommand{\ptblProjectMeanMetric}{(Table \emph{project mean metric}---18\,203\,053 records)}
\newcommand{\tblMostForkedInGroup}{\emph{most forked in group} (7\,553\,705 records)}
\newcommand{\ptblrecMostForkedInGroup}{(7\,553\,705 records)}
\newcommand{\ptblMostForkedInGroup}{(Table \emph{most forked in group}---7\,553\,705 records)}
\newcommand{\tblNoTable}{\emph{no table} (0 records)}
\newcommand{\ptblrecNoTable}{(0 records)}
\newcommand{\ptblNoTable}{(Table \emph{no table}---0 records)}
\newcommand{\tblDeduplicateByStars}{\emph{deduplicate by stars} (10\,649\,348 records)}
\newcommand{\ptblrecDeduplicateByStars}{(10\,649\,348 records)}
\newcommand{\ptblDeduplicateByStars}{(Table \emph{deduplicate by stars}---10\,649\,348 records)}
\newcommand{\tblNoiseProjects}{\emph{noise projects} (37\,333\,119 records)}
\newcommand{\ptblrecNoiseProjects}{(37\,333\,119 records)}
\newcommand{\ptblNoiseProjects}{(Table \emph{noise projects}---37\,333\,119 records)}
\newcommand{\tblMeanStarMismatch}{\emph{mean star mismatch} (534\,903 records)}
\newcommand{\ptblrecMeanStarMismatch}{(534\,903 records)}
\newcommand{\ptblMeanStarMismatch}{(Table \emph{mean star mismatch}---534\,903 records)}
\newcommand{\tblProjectMetrics}{\emph{project metrics} (18\,203\,053 records)}
\newcommand{\ptblrecProjectMetrics}{(18\,203\,053 records)}
\newcommand{\ptblProjectMetrics}{(Table \emph{project metrics}---18\,203\,053 records)}
\newcommand{\tblProjectForks}{\emph{project forks} (6\,958\,551 records)}
\newcommand{\ptblrecProjectForks}{(6\,958\,551 records)}
\newcommand{\ptblProjectForks}{(Table \emph{project forks}---6\,958\,551 records)}
\newcommand{\tblAcgroups}{\emph{acgroups} (18\,203\,053 records)}
\newcommand{\ptblrecAcgroups}{(18\,203\,053 records)}
\newcommand{\ptblAcgroups}{(Table \emph{acgroups}---18\,203\,053 records)}
\newcommand{\tblForksClonesNoise}{\emph{forks clones noise} (50\,324\,363 records)}
\newcommand{\ptblrecForksClonesNoise}{(50\,324\,363 records)}
\newcommand{\ptblForksClonesNoise}{(Table \emph{forks clones noise}---50\,324\,363 records)}
\newcommand{\tblProjectIssues}{\emph{project issues} (9\,498\,704 records)}
\newcommand{\ptblrecProjectIssues}{(9\,498\,704 records)}
\newcommand{\ptblProjectIssues}{(Table \emph{project issues}---9\,498\,704 records)}
\newcommand{\tblCommitPercentiles}{\emph{commit percentiles} (32 records)}
\newcommand{\ptblrecCommitPercentiles}{(32 records)}
\newcommand{\ptblCommitPercentiles}{(Table \emph{commit percentiles}---32 records)}
\newcommand{\tblDeduplicateByForks}{\emph{deduplicate by forks} (10\,649\,348 records)}
\newcommand{\ptblrecDeduplicateByForks}{(10\,649\,348 records)}
\newcommand{\ptblDeduplicateByForks}{(Table \emph{deduplicate by forks}---10\,649\,348 records)}

\fi

\ifx\extended\relax
\maketitle
\else

\ccsdesc[100]{Software and its engineering~Open source model}
\ccsdesc[300]{General and reference~Empirical studies}
\ccsdesc[500]{Software and its engineering~Software configuration management and version control systems}


\keywords{\kwlist}

\fi

\begin{abstract}
GitHub projects can be easily replicated through the site's fork process or
through a Git clone-push sequence.
This is a problem for empirical software engineering, because it can
lead to skewed results or mistrained machine learning models.
We provide a dataset of 10.6 million GitHub projects that are copies of others,
and link each record with the project's ultimate parent.
The ultimate parents were derived from a ranking along six metrics.
The related projects were calculated as the connected components
of an 18.2 million node and 12 million edge denoised graph
created by directing edges to ultimate parents.
The graph was created by filtering out
more than 30 hand-picked and 2.3 million pattern-matched clumping projects.
Projects that introduced unwanted clumping were identified
by repeatedly visualizing shortest path distances between unrelated
important projects.
Our dataset identified 30 thousand duplicate projects in an existing
popular reference dataset of 1.8 million projects.
An evaluation of our dataset against another created independently with
different methods found a significant overlap, but also differences
attributed to the operational definition of what projects are
considered as related.
\end{abstract}

\ifx\extended\relax

\noindent\textbf{Keywords:} \kwlist

\blfootnote{This is a technical note expanding reference~\cite{SKM20p},
which should be cited in preference to this text.}

\epigraph{In theory, there is no difference between theory and practice, while in practice, there is.}{Benjamin Brewster}

\else

\maketitle

\fi

\section{Introduction} 
\label{sec:intro}
Anyone can create a copy of a GitHub project through a single effortless
click on the project's {\em fork} button.
Similarly, one can also create a repository copy with just two Git commands.
Consequently, GitHub contains many millions of copied projects.
This is a problem for empirical software engineering.
First, when data containing multiple copies of a repository are analyzed,
the results can end up skewed~\cite{LMMS17}.
Second, when such data are used to train machine learning models,
the corresponding models can behave incorrectly~\cite{ID18,All19}.

In theory, it should be easy to filter away copied projects.
The project details provided by the GitHub API contain the field
{\tt fork}, which is {\em true} for forked projects.
They also include fields under {\tt parent} or {\tt source},
which contain data concerning the fork source.

In practice, the challenges of detecting and grouping together
copied GitHub repositories are formidable.
At the computational level, they involve finding among hundreds of
millions of projects those that are near in a space of billions of
dimensions (potentially shared commits).
The use of GitHub
for courses and coursework with hundreds of thousands of participants,\footnote{\url{https://github.com/rdpeng/ProgrammingAssignment2}}
for experimenting with version control systems,\footnote{\url{https://archive.softwareheritage.org/browse/search/?q=dvcsconnectortest&with_visit&with_content}}
and for all kinds of frivolous or mischievous activity\footnote{\url{https://github.com/illacceptanything/illacceptanything}}
further complicates matters.

\ifx\extended\relax
In the following sections we present
how we created a dataset identifying and grouping together
GitHub projects with shared ancestry or commits (Sections~\ref{sec:method}
and~\ref{sec:problems}),
the data schema and availability (Section~\ref{sec:overview}),
an evaluation of data quality (Section~\ref{sec:evaluation}),
indicative findings based on the dataset (Section~\ref{sec:findings}),
related work (Section~\ref{sec:related}), and
ideas for research and improvements (Section~\ref{sec:ideas}).
\fi

\section{Dataset Creation} 
\label{sec:method}
\ifx\extended\relax
\afterpage{
\begin{landscape} 
\begin{figure}
\includegraphics[width=1.4\textheight]{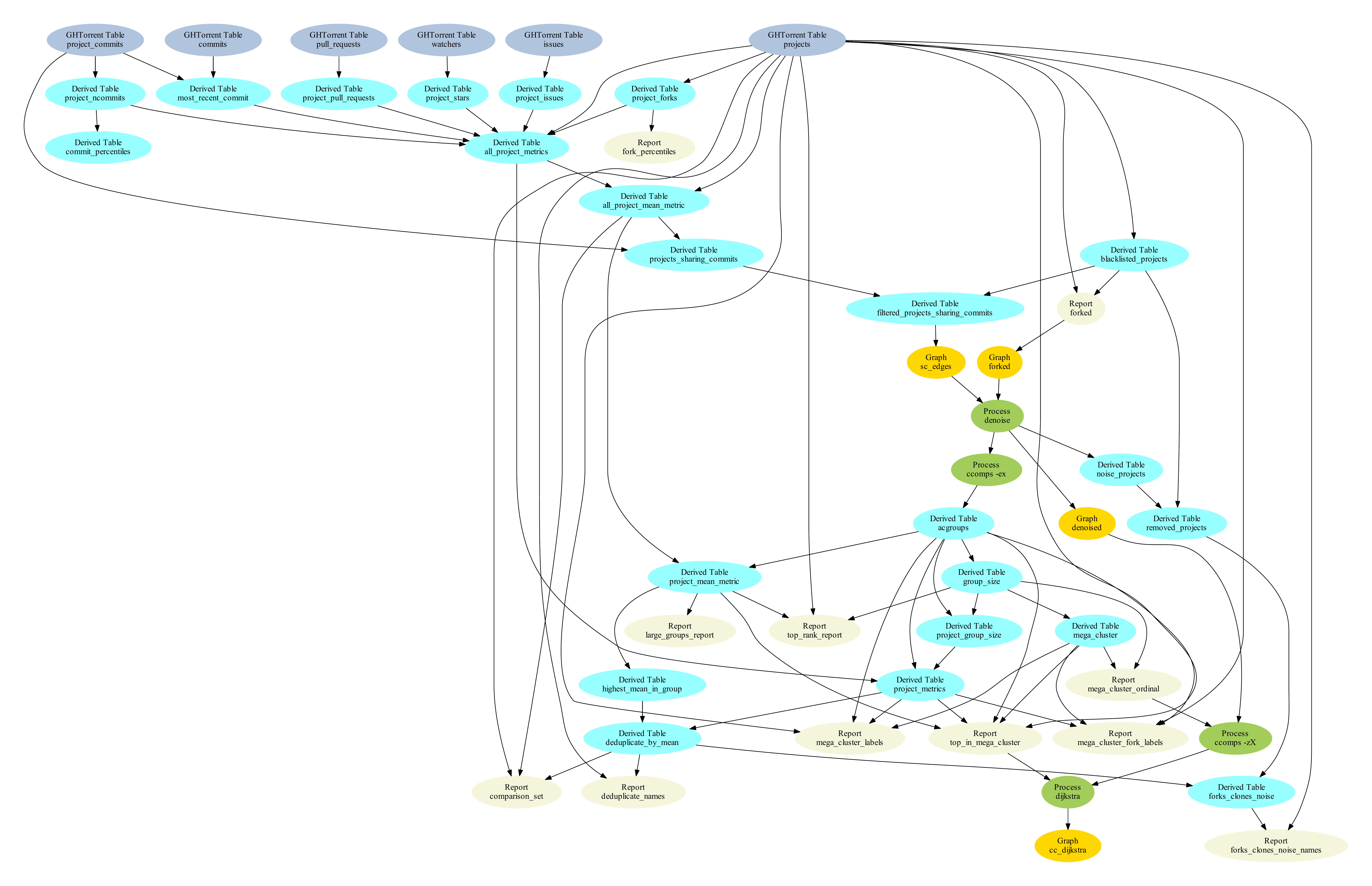}
\caption{Overview of the dataset creation process}
\label{fig:dataflow}
\end{figure}
\end{landscape}
}
\fi
An overview of the dataset's construction process is depicted in
\ifx\extended\relax
Figure~\ref{fig:dataflow}.
\else
an extended version of this paper~\cite{SKM20b}.
\fi
The projects were selected from GitHub by analyzing
the GHTorrent~\cite{GS12,Gou13} dataset (release 2019-06-01)
by means of the {\em simple-rolap} relational online analytical processing
and {\em rdbunit} relational unit testing frameworks~\cite{GS17}.
Following published recommendations~\cite{IHG12},
the code and primary data associated with this endeavor
are openly available online,\footnote{\url{https://doi.org/10.5281/zenodo.3742818}}
and can be used to replicate the dataset or
construct an updated version from newer data.


The GHTorrent dataset release we used contains
details about
125 million (125\,486\,232) projects,
one billion (1\,\-368\,\-235\,\-072) individual commits, and
six billion (6\,251\,898\,944) commits associated with
(possibly multiple, due to forks and merges) projects.

\ifx\extended\undefined
\begin{lstlisting}[language=sql,
float,
morekeywords={rank,partition,over,first_value,BIGINT},
caption={Identification of projects with common commits},
label={l:ccommits},
belowskip=-1.5\baselineskip
]
select distinct p1, p2 from (
  select project_commits.project_id as p2,
    first_value(project_commits.project_id) over (
      partition by commit_id
      order by mean_metric desc) as p1
    from project_commits
    inner join forkproj.all_project_mean_metric
    on all_project_mean_metric.project_id = project_commits.project_id) as shared_commits
where p1 != p2;
\end{lstlisting}
\fi

We first grouped shared commits to a single ``attractor'' project,
which was derived based on the geometric mean \ptblAllProjectMeanMetric{}
of six quality attributes:\footnote{In the interest of readability,
this text replaces the underscores in the table names with spaces.}
recency of the latest commit \ptblMostRecentCommit{},
as well as the
number of stars \ptblProjectStars{},
forks \ptblProjectForks{},
commits \ptblProjectNcommits{},
issues \ptblProjectIssues{},
and pull requests \ptblProjectPullRequests{}.
In addition, the project-id was used as a tie-breaker.
To avoid the information loss caused by zero values~\cite{Hab12},
we employed the following formula proposed by del Cruz and Kref~\cite{CK18}:
\[
G_{\epsilon, X}(X) =
\exp\left(
  \frac{1}{n}\sum_{i=1}^{n}{\log\left(x_i + \delta_{*}\right)}
\right)-\delta_{*}
\]
with $\delta_{*}$ calculated to have the value of 0.001 for our data.
We used a query utilizing the SQL window functions in order
to group together shared commits \ptblProjectsSharingCommits{}
without creating excessively large
result sets
\ifx\extended\relax
(see Listing~\ref{l:HighestMeanInGroup} in the appendix).
\else
(see Listing~\ref{l:ccommits}).
\fi

To cover holes in the coverage of shared commits,
we complemented the \emph{projects sharing commits} table
with projects related by GitHub project forks,
after removing a set of hand-picked and heuristic-derived
projects that mistakenly linked together
unrelated clusters \ptblBlacklistedProjects{}.
In addition, we removed from the combined graph
non-isolated nodes having between two and five edges, in order to reduce
unwanted merges between unrelated shared commit and fork clusters.
The method for creating the blacklisting table, along with the details
behind the denoising process, are explained in Section~\ref{sec:problems}.

We subsequently converted projects with shared commits or shared fork ancestry
into an (unweighted) graph to find its connected
components, which would be the groups of related projects.
We identified connected components using the {\em GraphViz}~\cite{GN00}
{\em ccomps} tool \ptblAcgroups{}.
The final steps involved
determining the size of each group \ptblGroupSize{},
associating it with each project \ptblProjectGroupSize{}
and its metrics \ptblProjectMetrics{},
obtaining the mean metric for the selected projects \ptblProjectMeanMetric{},
and associating with each group the project excelling in the metric
\ptblHighestMeanInGroup{}.
This was then used to create the deduplication table \ptblDeduplicateByMean{}
by partnering each project with a sibling having the highest mean value
in the calculated metrics.

\section{Down the Rabbit Hole} 
\label{sec:problems}
We arrived at the described process after numerous false starts, experiments,
and considerable manual effort.
\ifx\extended\relax
Some problems included
the discovery that in GHTorrent shared commits do not always lead to the
same ancestral commit,
a query that exhausted 128GB of RAM, and
a result set that exceeded the maximum number of rows supported by
PostgreSQL (four billion).
\fi
Here we provide details regarding the technical difficulties
associated with the dataset's creation,
the rationale for the design of the adopted processing pipeline, and
the process of the required hand-cleaning and denoising.

\ifx\extended\relax
Our initial plan for reducing the problem's high dimensionality involved
associating each commit with a parent-less ancestor.
In theory, each graph component of copied projects would have a unique such ancestor,
allowing us to easily group projects together.
In practice, we found that the commit history is incomplete, and that various
projects share multiple ancestries.

Our next approach for finding shared commits was based on
grouping the records of the project and commit identifier table
by the commit identifier, to identify projects with common commits.
This process ran out of memory on a 128 GB RAM machine.
We tried to run a query to export the data into a file,
but this also failed with an error
``PGresult cannot support more than INT\_MAX tuples''.
In the end, we resorted to dumping the commit and project identifier table
with the database's dump utility, {\em pg\_dump}, and filtering the output
to obtain the required data.
We sorted the file by the commit identifier as the primary key
(Unix sort can handle arbitrary amounts of data by sorting in batches
and then merge-sorting the intermediate files),
and then used a small {\em awk} script to create in memory (9 GB RAM)
a set of projects sharing commits,
thus reducing the amount of downstream data.
\fi

Our manual verification of large graph components
uncovered a mega-component of projects with 4\,278\,791 members.
Among the component's projects were many
seemingly unrelated popular ones,
such as the following ten:
FreeCodeCamp/FreeCodeCamp,
facebook/react,
getify/You-Dont-Know-JS,
robbyrussell/oh-my-zsh,
twbs/bootstrap,
Microsoft/vscode,
github/gitignore,
torvalds/linux,
nodejs/node, and
flutter/flutter.

\ifx\extended\relax
We experimented with introducing weights to the graph,
changing the edge-creation algorithm to include only projects
sharing at least two commits.
According to a percentile calculation of commits in GHTorrent,
this could have drastic consequences,
because only the top 60\% of projects ordered by the number of
recorded commits have more than two commits.
The size of the mega-component was reduced,
but it still contained 3\,202\,377 members.
(The number of components also fell from 6\,103\,690 to 6\,065\,658.)
Given that weights did not break up the mega-component,
we eventually kept the graph unweighted.
\fi


We wrote a graph-processing script to remove from the graph
all but one edges from projects with up to five edges.
We chose five based on the average number of edges per node with more
than one edge (3.8) increased by one for safety.
We also looked at the effect of other values.
Increasing the denoising limit up to ten edges reduced the size of
the mega-component only little to 2\,523\,841.
Consequently, we kept it at five to avoid removing too many duplicate
projects.
This improved somewhat the situation,
reducing the size of the mega-component to 2\,881\,473 members.

Studying the mega-component we observed that many attractor projects were
personal web sites.\graphref{popular-4683d07}\textsuperscript{,}\footnote{The
referenced graph images G\emph{N} are distributed
with the paper's replication package.
\ifx\extended\relax
All but one are also included in this note's appendix as zoomable figures.
The yellow-colored nodes are
the ones belonging in the set of top-ranked projects.
\fi
}
Focusing on them we found that apparently many
clone a particular personal style builder,
build on it,
force push the commits, and then
repeat the process with another builder project.
For example,
it seems that through this process,
wicky-info/wicky-info.github.io
shares commits with 139 other projects.

Based on this insight we excluded projects with names indicating
web sites (all ending in github.io), and also removed from the graph
nodes having between two and five edges, considering them as adding noise.
This reduced considerably the mega-component size in the graph of
projects with shared commits, to the point where the largest
component consisted mostly of programming assignments forked
and copied thousands of times
(jtleek/datasharing --- 199k forks,
rdpeng/ProgrammingAssignment2 --- 119k,
rdpeng/RepData\_PeerAssessment1 --- 32.3k).

We also joined the generation of the fork tree and the common commit graph
to reduce their interference,
applying the denoising algorithm to both.
This reduced the clusters to very reasonable sizes,
breaking the mega-component to only include a few unrelated
projects,\graphref{popular-ff65bf7}
which was further improved by blacklisting a couple of
Android Open Source Project repositories.\graphref{popular-21bfc4e}

We manually inspected the five projects
with the highest mean ranking in each of the first 250 clusters,
which comprise about 1.6 million projects.
The most populous component (Linux) had 175\,184 members and
the last, least populous, component (vim) had 1912 members.
Many cases of several high-ranked projects in the same component
involved genuine forks.
This is for example the case of
MariaDB/server linking percona/percona-server, mysql/mysql-server, and
facebook/mysql-8.0, among others.\graphref{top-3c639f2-mariadb}
Where these referred to different projects,
we drew a map of shortest path between the 49 top-ranked
projects and the first or 50th one,
and blacklisted low-ranked projects that were linking together
unrelated repositories.

Resolved examples include the linking of
Docker with Go,\graphref{top-d516729-docker}
Django with Ruby on Rails,\graphref{top-d516729-django-rails}
Google projects with zlib,\graphref{top-d516729-Google-zlib}
Diaspora with Arduino,\graphref{top-d516729-Diaspora-Arduino}
Elastic Search with Pandas,\graphref{top-d516729-Elastic-Pandas}
Definitely Types with RxJS,\graphref{top-d516729-Definitely-Typed-RxJS}
Ansible with Puppet,\graphref{top-d516729-Ansible-Puppet}
PantomJS with WebKit and Qt,\graphref{top-d516729-qt}
OpenStack projects,\graphref{top-d516729-openstack}
Puppet modules,\graphref{top-d516729-puppet-modules}
documentation projects,\graphref{top-d516729-docs}
Docker registry with others,\graphref{top-d516729-archived}
Drupal with Backdrop,\graphref{top-d516729-backgrop-drupal}
Python and Clojure koans,\graphref{top-3c639f2-python-clojure-koans}
Vimium with Hubot,\graphref{top-3c639f2-vimium-hubot}
as well as several ASP.NET projects.\graphref{top-d516729-aspnet}

For some clusters that failed to break up we repeated the exercise,
looking at paths in the opposite direction,
removing additional projects such as those linking
Linux with Dagger,\graphref{top-2b2b176-linux}
Ruby with JRuby, oh-my-zsh, Capistrano,
and git-scm,\graphref{top-2b2b176-capistrano} and
Laravel with Fuel.\graphref{top-2b2b176-laravel-fuel}

In some cases the culprits were high-ranked projects, such as
boostorg/spirit, which links together more than ten Boost
repositories,\graphref{top-d516729-Google-zlib}
apache/hadoop, which links with Intel-bigdata/SSM,\graphref{top-d516729-hadoop-ssm}
DefinitelyTyped/DefinitelyTyped,
which links to Reactive-\-Extensions/\-RxJS,\graphref{top-d516729-Definitely-Typed-RxJS}
ReactiveX/RxJava, which links several Netflix repositories,\graphref{top-d516729-RcJava-Netflix}
jashkenas/underscore, which links with lodash/lodash,\graphref{top-d516729-lodash-underscore}
jsbin/jsbin linking to cdnjs/cdnjs,\graphref{top-d516729-cdnjs-}
ravendb/ravendb linking to SignalR/SignalR,\graphref{top-d516729-signalr-ravendb}
Kibana linked with Grafana,\graphref{top-d516729-kibana-grafana}
CartoDB/carto linked with less/\-less.js.\graphref{top-3c639f2-less-carto}
Other projects, such as Swift and LLVM,\graphref{top-481f449-Swift-LLVM}
or Docker with Con\-tai\-nerd,\graphref{top-2b2b176-docker-containerd}
were too entangled to bring apart.

\ifx\extended\relax
We also looked at the number of edges of each node in the component,
reasoning that a single project was somehow acting as a hub,
gluing all disparate projects together.
We indeed found a project (jlord/patchwork) with a very large number
of edges (11\,735), but these were consistent with the number of its forks
(31\,449), and also connections between unrelated projects were not passing
through it.
\fi

To further investigate what brings the component's projects together,
we selected from the component one popular project with relatively few
forks (creationix/nvm), and applied Dijkstra's shortest path algorithm to
find how other projects got connected to it.
We drew paths from that project to 30 other popular projects belonging to
the same component, and started verifying each one by hand.
We looked at the shared commits between unrelated projects that we
found connected, such as yui-knk/rails and seuros/django.

Some (very few) commits appear to be shared by an inordinate number of projects.
At the top,
three commits are shared by 100\,683 projects,
another three by 67\,280, and
then four by 53\,312.
However, these numbers are not necessarily wrong, because there
are five projects with a correspondingly large number of forks:
125\,491 (jtleek/datasharing),
124\,326 (rdpeng/ProgrammingAssignment2),
111\,986 (octocat/Spoon-Knife),
70\,137 (tensorflow/tensorflow), and
66\,066 (twbs/bootstrap).
The first two commits are associated with many (now defunct) projects
of the user dvcsconnectortest
(missingcommitsfixproof, missingcommitstest, and then
missingcommitstest\_250\_1393252414399)
for many different trailing numbers.
However, the particular user is associated with very few commits,
namely 1240,
so it is unlikely that these commits have poisoned other components
through transitive closure.

\ifx\extended\relax
\begin{lstlisting}[language=awk,
float,
caption={Final implementation of denoising algorithm},
morekeywords={openF, N, E, ARGV, degreeOf, edge\_t, fstedge, nxtedge, opp},
label={l:denoise},
belowskip=-1.5\baselineskip
]
#!/usr/bin/gvpr -cf
#
# Remove nodes having between 2 and N edges writing their names
# to the file reports/noise_projects.txt.
# The removal reduces unwanted merges between shared commit and
# fork clusters.
#
# N is specified as an argument with -a
# gvpr -f denoise.gvpr -a 5
#

BEGIN {
  int nf = openF("reports/noise_projects.txt", "w");
  int noise_ceiling = ARGV[0];
}

N {
  int d = degreeOf($G, $);
  if (d > 1 && d <= noise_ceiling) {
      /*
       * Iterate through edges to sum the degrees of their nodes
       * If this equals the number of edges, then this clique is
       * disjoined and cannot join together other cliques.
       */
      int degree_sum = 0;
      edge_t e;
      for (e = fstedge($); e; e = nxtedge(e, $))
        degree_sum += degreeOf($G, opp(e, $));
      if (degree_sum > d) {
        printf(nf, "%s\n", $.name);
        delete($G, $);
      }
  }
}
\end{lstlisting}
\fi

We later on improved the denoising to incorporate components that could
be trivially determined for isolation,
by looking at just the neighboring nodes.
The algorithm we employed is applied to all nodes $n$ having between
two and five edges; the ones we used to consider as noise.
It sums up as $s$ being the number of edges of all nodes $n'$ that were directly
connected to $n$.
If $s$ is equal to the edges leading to $n$, then $n$ and its immediate
neighbors form a component,
otherwise it is considered as adding noise and is disconnected from
its neighbors.
For a graph with edges $E$ the condition for a node $n$ being considered
as noise, can be formally described as
\[
\left| (n, n') | (n, n') \in E \right|
\neq
\sum_{\forall n' | (n, n') \in E} {\left| \{(n', n'') | (n', n'') \in E\} \right| }
\]
%
Applying this algorithm
\ifx\extended\relax
(Listing~\ref{l:denoise})
\fi
decreased the number of ignored ``noise projects'' marginally
from 37\,660\,040 to 37\,333\,119,
increasing, as expected, the number of components by the same amount,
from  2\,145\,837 to 2\,472\,758,
and also increasing the number of projects considered as clones
by about double that amount,
from 9\,879\,677 to 10\,649\,348.

\section{Dataset Overview} 
\label{sec:overview}
The dataset is provided\footnote{\url{https://doi.org/10.5281/zenodo.3653920}}
as two files identifying GitHub repositories
using the {\em login-name}/{\em project-name} convention.
The file \emph{deduplicate\_names} contains 10\,649\,348 tab-separated
records mapping a duplicated {\em source project} to a definitive
{\em target project}.
The file \emph{forks\_clones\_noise\_names} is a 50\,324\,363 member
superset of the source projects, containing also projects that
were excluded from the mapping as noise.

The files are to be used as follows.
After selecting some projects for conducting an
empirical software engineering study with GitHub projects,
the first file should be used to map potentially duplicate projects
into a set of definitive ones.
Then, any remaining projects that appear in the second file should be
removed as these are likely to be low-value projects with a high probability
of undesirable duplication.

\section{Duplication in Existing Datasets} 
\label{sec:findings}
As an example of use of our dataset,
we deduplicated the Reaper dataset~\cite{MKCN17},
which contains scores concerning seven software engineering practices
for about 1.8 million (1\,853\,205) GitHub projects.
The study has influenced various subsequent works~\cite{RPW19,BD19,ARKS18,BONB18}
through the provided recommendations and filtering criteria
for curating collected repositories.
The authors have excluded deleted and forked projects,
considering the latter as near duplicates.

Around 30 thousand (30\,095) duplicate projects
were identified in the Reaper dataset using \emph{deduplicate\_names}.
\ifx\extended\relax
The first ten components with the most recurrences
involve the following ultimate parents and repetitions:
torvalds/linux (2614), gatsbyjs/gatsby (545),
boxen/our-boxen (229), backdrop/backdrop (121),
publify/publify (110), boostorg/boost (109),
llvm-mirror/llvm (108), laravel/framework (98),
universal-ctags/ctags (89), saasbook/hw3\_rottenpotatoes (87).
In addition, the deduplication of the 800 hand-picked projects
used in the classifiers' training and validation processes unveiled
only nine duplicate instances in the organization dataset,
one in the utility,
and none in the validation.
These include aspnet/Mvc (7), apache/flink (2),
mozilla/bedrock (2), and bitcoin/bitcoin (2) in the organization,
and torvalds/linux (2) in the utility.
\else
The deduplication of the 800 hand-picked projects
used in the classifiers' training and validation processes
unveiled ten duplicate instances.
\fi
Further investigation is required to measure any potential impact
of the ten duplicate projects on the classification outcome.
Nevertheless, researchers selecting projects from Reaper for their work
can benefit from our dataset to filter out duplicate occurrences,
to further improve the quality of selected projects
and avoid the problems outlined in Section~\ref{sec:intro}.

\section{Evaluation} 
\label{sec:evaluation}


\begin{table}
\caption{Dataset Comparison}
\label{tab:compare}
\ifx\extended\undefined
\vspace{-1ex}
\fi
\begin{center}
\begin{tabular}{lrr}
\hline
				& \multicolumn{2}{c}{Dataset} 	\\
Metric				&	CCFSC	& CDSC		\\
\hline
Number of repositories		& 10\,649\,348	& 116\,265\,607	\\
Number of independent projects	& 2\,470\,126	& 63\,829\,733	\\
Size of largest cluster		& 174\,919	& 244\,707	\\
Average cluster size		& 4.3		& 1.8		\\
Cluster size standard deviation & 169		& 44		\\
Reaper duplicates		& 30\,095	& 80\,079	\\
\hline
\end{tabular}
\end{center}
\vspace{-2ex}
\end{table}

We evaluated this dataset,
which was constructed by identifying connected components based
on forks and shared commits (CCFSC),
through a quantitative and qualitative comparison
with a similar dataset constructed using community detection of
shared commits (CDSC)~\cite{MKSD20}.
An overview of the basic characteristics of the two datasets appears in
Table~\ref{tab:compare}.
The two datasets share a substantial overlap both in terms
of source projects (8\,157\,317) and in terms of
cluster leaders (5\,513\,580).
On the other hand,
it is clear that CDSC is considerably more comprehensive than CCFSC
in order of magnitude,
covering more repositories.
An important factor in its favor is that it covers other forges
apart from GitHub, and therefore its population is a superset of CCFSC's.
However, if one also considers the projects that CCFSC considers as
noise (personal projects or projects with conflicting affiliations),
the overlap swells to 40\,338\,421, covering about a third of the total.
Furthermore, the fact that the increase in the Reaper dataset duplication
in the CDSC dataset is only about double that of the CCFSC dataset
indicates that the increased coverage of CCFSC may not be relevant
for some empirical software engineering studies.
These factors validate to some extent the dataset's composition.

To get a better understanding of where and how the two
datasets vary, we also performed a qualitative evaluation.
For this we selected a subgraph induced by the 1000 projects
with the highest geometric mean score,
and visualized the common and non-common elements of the 431
clusters that contained different nodes.\graphref{disagreements}
In 301 cases the clusters shared at least one common element.
The patterns we encountered mainly concern the following cases:
CCFSC links more (and irrelevant) clusters
compared to CDSC (e.g. FreeCodeCamp/FreeCodeCamp, gatsbyjs/gatsby,
robbyrussell/oh-my-zsh);
the converse happens (e.g. leveldb);
CCFSC clusters related projects that CDSC does not cluster
(e.g. tgstation/tgstation, bitcoin/bitcoin);
the converse happens (e.g. hdl\_qfs, t-s/blex);
there is considerable agreement between the two (e.g.
Homebrew/homebrew-core with afb/brew);
there is considerable agreement but
CDSC includes more related projects (e.g. aspnet/Mvc with h2h/Mvc).
In general, we noticed that CDSC appears to be more precise
at clustering than CCFSC, but worse at naming the clusters.

\section{Related Work} 
\label{sec:related}
In distributed version control and source code management platforms,
such as GitHub,
developers usually collaborate
using the pull request development model~\cite{GZ14,GPD14,GSB16,GZSD15},
according to which repositories are divided
into base and forked~\cite{KGBS16}.
This constitutes one of the perils of mining GitHub:
a repository is not necessarily a project~\cite{KGBS16},
with commits potentially differing between the associated repositories.

Code duplication in GitHub was studied by Lopes et al.~\cite{LMMS17}
through file-level and inter-project analysis
of a 4.5 million corpus of non-forked projects.
The overlap of files between projects,
as given by the files' token hashes,
was computed for certain thresholds and programming languages.
JavaScript prevails with 48\% of projects having at least 50\%
of files duplicated in other projects,
and 15\% of projects being 100\% duplicated.
Project-level duplication includes
appropriations that could be addressed by Git submodules,
abandoned derivative development,
forks with additional non-source code content,
and unorthodox uses of GitHub,
such as unpushed changes.
Code duplication can hamper the statistical reasoning
in random selections of projects,
and skew the conclusions of studies performed on them,
because the observations (projects) are not independent,
and diversity may be compromised.
For the converse problem of obtaining similar GitHub repositories
see the recent work by Phuong Nguyen and his colleagues~\cite{NRRD20}
and the references therein.

While it is common sense to select a sample
that is representative of a population,
the importance of diversity is often overlooked,
yet as important~\cite{All04}.
Especially in software engineering,
where processes of empirical studies often depend
on a large number of relevant context variables,
general conclusions are difficult to extract~\cite{BSL99}.
According to Nagappan et al.~\cite{NZB13},
to provide a good sample coverage,
selected projects should be diverse
rather than similar to each other.
Meanwhile,
increasing the sample size does not necessarily increase generality
when projects are not carefully selected.

Markovtsev and Kant in their work regarding topic modeling
of public repositories using names in source code~\cite{MK17},
recognized that
duplicate projects contain few original changes
and may introduce noise into the overall names distribution.
To exclude them and accelerate the training time of the topic model,
they applied Locality Sensitive Hashing~\cite{LRU20}
on the bag-of-words model.
According to the analysis,
duplicate repositories usually involve web sites,
such as github.io,
blogs and Linux-based firmwares,
which align with our observations.

A duplication issue was also identified
by Irolla and Dey~\cite{ID18} in the Drebin dataset~\cite{ASGR14},
which is often used
to assess the performance of malware detectors~\cite{PCYJ17,GMPB17}
and classifiers~\cite{GPMB16,SRKC16}.
Half of the samples in the dataset have other duplicate repackaged versions
of the same sequence of opcode.
Consequently,
a major part of the testing set may also be found in the training,
inflating the performance of the designed algorithms.
Experiments on classification algorithms
trained on the Drebin dataset
by including and excluding duplicates
suggested moderate to strong underrated inaccuracy,
and variation in the performance of the algorithms.

Similarly,
Allamanis examined the adverse effects of code duplication
in machine learning models of code~\cite{All19}. 
By comparing models trained
on duplicated and deduplicated code corpora,
Allamanis concluded that performance metrics,
from a user's perspective,
may be up to 100\% inflated
when duplicates are included.
The issue mainly applies to
code completion~\cite{RVY14,MT14},
type prediction~\cite{HBBA18,RVK15} and
code summarization~\cite{IKCZ16,APS16},
where models provide recommendations on new and unseen code.

\section{Research and Improvement Ideas} 
\label{sec:ideas}
The main purpose of the presented dataset is to improve the quality of
GitHub project samples that are used to conduct
empirical software engineering studies.
It would be interesting to see how such duplication affects published
results by replicating existing studies after deduplicating the projects
by means of this dataset.
In addition, the dataset can be used for investigating the ecosystem of
duplicated projects in terms of
activity,
duplication methods (forks vs commit pushes),
tree depth,
currency, or
trustworthiness.

The dataset can be further improved by including projects from other forges
and by applying more sophisticated cleaning algorithms.

\ifx\extended\relax
\subsection*{Acknowledgements}
\else
\begin{acks}
\fi
This work has received funding from the
European Union's Horizon 2020 research
and innovation programme under grant
agreement No. 825328.

\ifx\extended\relax
D. Spinellis created the dataset and its unit tests.
Z. Kotti evaluated the dataset and researched related work.
Both authors contributed equally to the paper's writing.
A. Mockus contributed the comparison dataset.
\else
\end{acks}
\fi

\bibliographystyle{ACM-Reference-Format}
\ifx\mypubs\relax
\bibliography{macro,IEEEabrv,ddspubs,myart,classics,coderead,unix,various,mybooks,bigdata,ieeestd,isostd}
\else
\bibliography{forkproj,dds}
\fi 

\ifx\extended\relax
\appendix
\section{Appendix: Key SQL Queries and Representative Graphs}
\pdfsuppresswarningpagegroup=1

\needspace{5\baselineskip}
\begin{lstlisting}[language=sql,
basicstyle=\small\ttfamily,
morekeywords={rank,partition,over,first_value,BIGINT},
captionpos=t,
abovecaptionskip=4ex,
caption={SQL query for deriving the table \emph{all project mean metric}},
label={l:AllProjectMeanMetric},
]
-- Mean metric of each project

create table forkproj.all_project_mean_metric AS
  select project_id,
    -- Geometric mean of (value + 0.001) ^ 5
    -- See delta.py for the calculation of 0.001
    (stars + .001) *
    (forks + .001) *
    (commits + .001) *
    (issues + .001) *
    (recency + .001) *
    (pull_requests + .001) *
    -- Tie breaker with a value lower than 1, based on project_id
    -- Earlier projects score higher
    (.1 - project_id / 10. / (select max(id) from projects)) as mean_metric
  from forkproj.all_project_metrics;

create index on forkproj.all_project_mean_metric(project_id);
\end{lstlisting}

\needspace{5\baselineskip}
\begin{lstlisting}[language=sql,
basicstyle=\small\ttfamily,
morekeywords={rank,partition,over,first_value,BIGINT},
captionpos=t,
abovecaptionskip=4ex,
caption={SQL query for deriving the table \emph{most recent commit}},
label={l:MostRecentCommit},
]
-- The most recent commit for each project

CREATE TABLE forkproj.most_recent_commit  AS
  select project_commits.project_id as project_id,
    max(created_at) as most_recent
  from commits
  inner join project_commits
  on project_commits.commit_id = commits.id
  group by project_commits.project_id;

create unique index on forkproj.most_recent_commit(project_id);
\end{lstlisting}

\needspace{5\baselineskip}
\begin{lstlisting}[language=sql,
basicstyle=\small\ttfamily,
morekeywords={rank,partition,over,first_value,BIGINT},
captionpos=t,
abovecaptionskip=4ex,
caption={SQL query for deriving the table \emph{project stars}},
label={l:ProjectStars},
]
-- Number of stars per project

create table forkproj.project_stars AS
  select repo_id as id, count(*) as stars
  from watchers
  group by repo_id;

create index on forkproj.project_stars(id);
\end{lstlisting}

\needspace{5\baselineskip}
\begin{lstlisting}[language=sql,
basicstyle=\small\ttfamily,
morekeywords={rank,partition,over,first_value,BIGINT},
captionpos=t,
abovecaptionskip=4ex,
caption={SQL query for deriving the table \emph{project forks}},
label={l:ProjectForks},
]
-- Number of forks per project

create table forkproj.project_forks  AS
  select
    projects.forked_from as id,
    count(*) as forks
  from projects
  where forked_from is not null
  group by forked_from;

create index on forkproj.project_forks(id);
\end{lstlisting}

\needspace{5\baselineskip}
\begin{lstlisting}[language=sql,
basicstyle=\small\ttfamily,
morekeywords={rank,partition,over,first_value,BIGINT},
captionpos=t,
abovecaptionskip=4ex,
caption={SQL query for deriving the table \emph{project ncommits}},
label={l:ProjectNcommits},
]
-- Number of commits per project

create table forkproj.project_ncommits AS
  select project_id as id, count(project_id) as commits
  from project_commits
  group by project_id;

create index on forkproj.project_ncommits(id);
\end{lstlisting}

\needspace{5\baselineskip}
\begin{lstlisting}[language=sql,
basicstyle=\small\ttfamily,
morekeywords={rank,partition,over,first_value,BIGINT},
captionpos=t,
abovecaptionskip=4ex,
caption={SQL query for deriving the table \emph{project issues}},
label={l:ProjectIssues},
]
-- Number of project issues per project

create table forkproj.project_issues AS
  select repo_id as id, count(*) as issues
  from issues
  group by repo_id;

create index on forkproj.project_issues(id);
\end{lstlisting}

\needspace{5\baselineskip}
\begin{lstlisting}[language=sql,
basicstyle=\small\ttfamily,
morekeywords={rank,partition,over,first_value,BIGINT},
captionpos=t,
abovecaptionskip=4ex,
caption={SQL query for deriving the table \emph{project pull requests}},
label={l:ProjectPullRequests},
]
-- Number of pull requests per project

create table forkproj.project_pull_requests AS
  select base_repo_id as id, count(*) as pull_requests
  from pull_requests
  group by base_repo_id;

create index on forkproj.project_pull_requests(id);
\end{lstlisting}

\needspace{5\baselineskip}
\begin{lstlisting}[language=sql,
basicstyle=\small\ttfamily,
morekeywords={rank,partition,over,first_value,BIGINT},
captionpos=t,
abovecaptionskip=4ex,
caption={SQL query for deriving the table \emph{projects sharing commits}},
label={l:ProjectsSharingCommits},
]
-- Projects sharing commits with linked to the project with
-- the highest metric
-- metric(p1) > metric(p2)

create table forkproj.projects_sharing_commits as
  select distinct p1, p2 from (
    select project_commits.project_id as p2,
      first_value(project_commits.project_id) over (
	partition by commit_id
	-- Link all with the oldest project
        order by mean_metric desc) as p1
      from project_commits
      inner join forkproj.all_project_mean_metric
      on all_project_mean_metric.project_id = project_commits.project_id
  ) as shared_commits
  where p1 != p2;
\end{lstlisting}

\needspace{5\baselineskip}
\begin{lstlisting}[language=sql,
basicstyle=\small\ttfamily,
morekeywords={rank,partition,over,first_value,BIGINT},
captionpos=t,
abovecaptionskip=4ex,
caption={SQL query for deriving the table \emph{blacklisted projects}},
label={l:BlacklistedProjects},
]
-- Project ids of personal sites

create table forkproj.blacklisted_projects AS
  select id from projects where name like '%.github.io' or
    substr(url, 30) in (
      'illacceptanything/illacceptanything',
      'github/gitignore',
      'android/platform_build',		-- Links to Dagger and Simple Gallery
      'Reese-D/my_emacs',		-- Links oh-my-zsh, flydeck, magit, ...
      'destructuring/junas',		-- Links capistrano, janus, git-scm
      'yosadchuk/git-scm.com',		-- Links git-scm with progit
      'carsomyr/rbenv-ubuntu',		-- Links ruby, jruy, pyenv
      'expl0ratory/.vim',		-- Links oh-my-zsh, vim-airline, ...
      'scrooloose/syntastic',		-- Links YouCompleteMe, ...
      'jbarros/checkbook',		-- Links Laravel with Fuel
      'bogner/llvm-zipper-prototype',	-- LLVM monorepo with 612k commits, joining swift, klee, LLVM
      'TurboROM/MERGE_test',		-- Deleted; brings flutter
      'AdrianDC/aosp_development_sony8960_o_mr1', -- Archived; join v8 with other C
      'AdrianDC/aosp_development_sony8960_p', -- See above
      'LineageOS/android_packages_apps_WallpaperPicker', -- 2 stars; joins LLVM w. Android
      'cmars/tools',			-- Links Docker with glide, vcs
      'cmars/oo',			-- Links Docker with golang/*
      'seuros/django',			-- Links Django with Rails
      'phantom9999/folly',		-- Links diverse Google projects, zlib, ...
      'chriswingler/Simulator',		-- Links diaspora with Arduino
      'gfyoung/elasticsearch',		-- Links elasticsearch with Pandas
      'Inuits/ensible',			-- Links ansible with Puppet
      'chapuni/llvm-project',		-- Joins Swift and Klee
      'dotnet-maestro-bot/Common',	-- Joins several aspnet projects
      'dotnet-maestro-bot/AspNetCore',	-- See above
      'lodejard/AllNetCore',		-- See above
      'Vitallium/phantomjs-qt5',	-- PhantomJS, Qt, WebKit
      'derekhiggins/delorean-specs',	-- Joins several OpenStack projects
      'kscherer/puppet-modules',	-- Join Puppet modules
      'akeif/compucorp-task1-envs',	-- Join Puppet modules
      'harikt/docs',			-- Joins several documentations
      'saltlakeryan/archived-projects',	-- Docker-registry, microHTTPD, ...
      'jenlampton/badcamp2014',		-- Backdrop with Drupal
      'tokyo-jesus/university',		-- PYthon and clojure koans
      'decaffeinate-examples/vimium',	-- Links vimium with hubot
      'octocat/Spoon-Knife');		-- Sever hundred k forks for training

create index on forkproj.blacklisted_projects(id);
\end{lstlisting}

\needspace{5\baselineskip}
\begin{lstlisting}[language=sql,
basicstyle=\small\ttfamily,
morekeywords={rank,partition,over,first_value,BIGINT},
captionpos=t,
abovecaptionskip=4ex,
caption={SQL query for deriving the table \emph{acgroups}},
label={l:Acgroups},
]
-- Import connected components of all forked projects

create table forkproj.acgroups (
  cc_id BIGINT not null,
  project_id BIGINT not null,
  primary key(project_id, cc_id)
);

\copy forkproj.acgroups from 'reports/ac_groups.csv' CSV;

create unique index on forkproj.acgroups(project_id);
create index on forkproj.acgroups(cc_id);
\end{lstlisting}

\needspace{5\baselineskip}
\begin{lstlisting}[language=sql,
basicstyle=\small\ttfamily,
morekeywords={rank,partition,over,first_value,BIGINT},
captionpos=t,
abovecaptionskip=4ex,
caption={SQL query for deriving the table \emph{group size}},
label={l:GroupSize},
]
-- Number of group members per group

create table forkproj.group_size AS
  select cc_id as id, count(cc_id) as n_group_members
  from forkproj.acgroups
  group by cc_id
  having count(*) > 1;

create index on forkproj.group_size(id);
\end{lstlisting}

\needspace{5\baselineskip}
\begin{lstlisting}[language=sql,
basicstyle=\small\ttfamily,
morekeywords={rank,partition,over,first_value,BIGINT},
captionpos=t,
abovecaptionskip=4ex,
caption={SQL query for deriving the table \emph{project group size}},
label={l:ProjectGroupSize},
]
-- Number of members in each project's group

create table forkproj.project_group_size AS
  select forkproj.acgroups.project_id as id, n_group_members
  from forkproj.acgroups
  left join forkproj.group_size
  on group_size.id = acgroups.cc_id;

create index on forkproj.project_group_size(id);
\end{lstlisting}

\needspace{5\baselineskip}
\begin{lstlisting}[language=sql,
basicstyle=\small\ttfamily,
morekeywords={rank,partition,over,first_value,BIGINT},
captionpos=t,
abovecaptionskip=4ex,
caption={SQL query for deriving the table \emph{project metrics}},
label={l:ProjectMetrics},
]
-- Number of stars, forks, commits, group members per project with clones

create table forkproj.project_metrics AS
  select forkproj.acgroups.project_id,
    forkproj.acgroups.cc_id as group_id,
    all_project_metrics.stars,
    all_project_metrics.forks,
    all_project_metrics.commits,
    all_project_metrics.issues,
    all_project_metrics.recency,
    all_project_metrics.pull_requests,
    n_group_members
  from forkproj.acgroups
  inner join forkproj.all_project_metrics
  on forkproj.all_project_metrics.project_id = forkproj.acgroups.project_id
  left join forkproj.project_group_size
  on forkproj.project_group_size.id = forkproj.acgroups.project_id;

create index on forkproj.project_metrics(project_id);
create index on forkproj.project_metrics(group_id);
\end{lstlisting}

\needspace{5\baselineskip}
\begin{lstlisting}[language=sql,
basicstyle=\small\ttfamily,
morekeywords={rank,partition,over,first_value,BIGINT},
captionpos=t,
abovecaptionskip=4ex,
caption={SQL query for deriving the table \emph{project mean metric}},
label={l:ProjectMeanMetric},
]
-- Project mean metric for each project in the group

create table forkproj.project_mean_metric AS
  select cc_id as group_id, acgroups.project_id, mean_metric
    from forkproj.acgroups
    inner join forkproj.all_project_mean_metric
    on forkproj.all_project_mean_metric.project_id = forkproj.acgroups.project_id;

create index on forkproj.project_mean_metric(project_id);
\end{lstlisting}

\needspace{5\baselineskip}
\begin{lstlisting}[language=sql,
basicstyle=\small\ttfamily,
morekeywords={rank,partition,over,first_value,BIGINT},
captionpos=t,
abovecaptionskip=4ex,
caption={SQL query for deriving the table \emph{highest mean in group}},
label={l:HighestMeanInGroup},
]
-- Project for each group excelling in its metrics

create table forkproj.highest_mean_in_group AS
  select group_id, project_id from (
      select project_id, group_id,
      rank() over (partition by group_id
	order by mean_metric desc) as mean_rank
      from forkproj.project_mean_metric
    ) as ranked
  where mean_rank = 1;

create index on forkproj.highest_mean_in_group(group_id);
\end{lstlisting}

\needspace{5\baselineskip}
\begin{lstlisting}[language=sql,
basicstyle=\small\ttfamily,
morekeywords={rank,partition,over,first_value,BIGINT},
captionpos=t,
abovecaptionskip=4ex,
caption={SQL query for deriving the table \emph{deduplicate by mean}},
label={l:DeduplicateByMean},
]
-- Partner each project with the sibling with the highest mean value
-- in the calculated metrics.
-- Each source project is to be mapped to the corresponding target.

create table forkproj.deduplicate_by_mean AS
  select project_metrics.project_id as source_id,
    highest_mean_in_group.project_id as target_id
  from forkproj.project_metrics
  left join forkproj.highest_mean_in_group
    on highest_mean_in_group .group_id = project_metrics.group_id
  where project_metrics.project_id != highest_mean_in_group.project_id;

create index on forkproj.deduplicate_by_mean(source_id);
\end{lstlisting}
\maxdeadcycles=1000
\begin{figure*}[h]
\includegraphics[width=\textwidth]{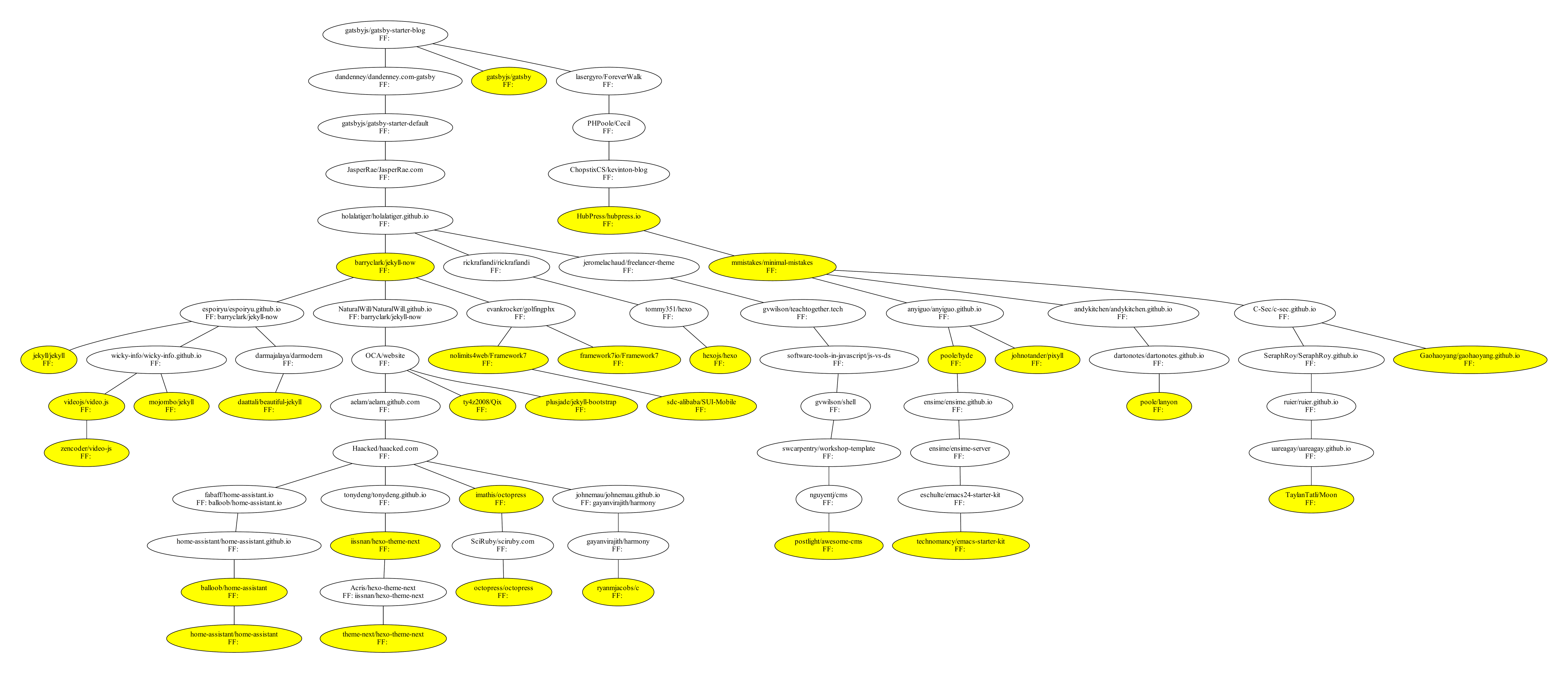}
\caption{Links associated with popular projects}
\label{fig:popular-4683d07}
\vspace{-1ex}
\end{figure*}
\begin{figure*}[h]
\includegraphics[width=\textwidth]{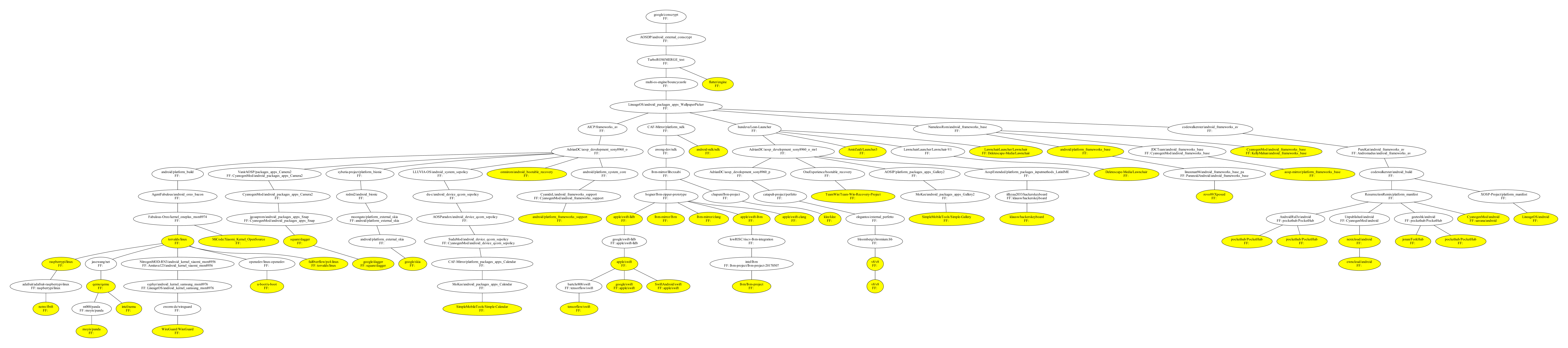}
\caption{Links associated with popular projects}
\label{fig:popular-ff65bf7}
\vspace{-1ex}
\end{figure*}
\begin{figure*}[h]
\includegraphics[width=\textwidth]{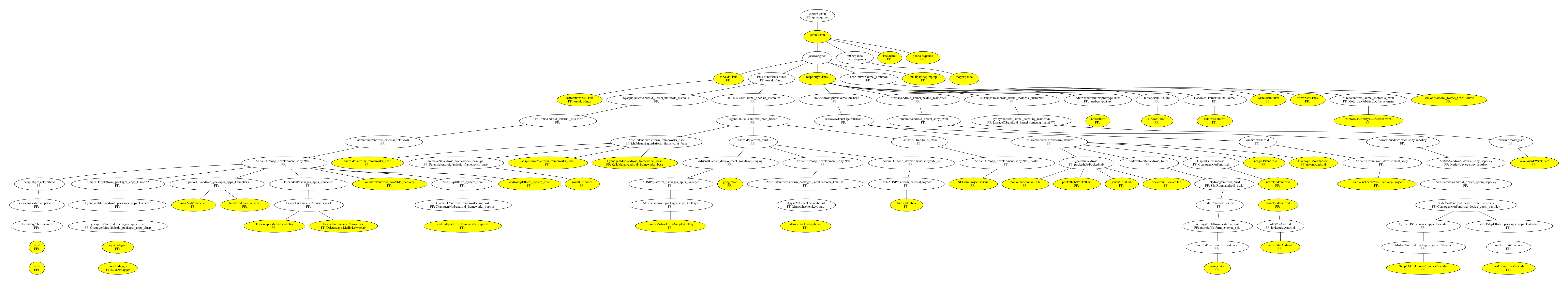}
\caption{Links associated with popular projects}
\label{fig:popular-21bfc4e}
\vspace{-1ex}
\end{figure*}
\begin{figure*}[h]
\includegraphics[width=\textwidth]{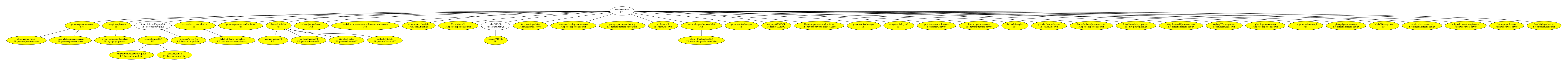}
\caption{Links associated with mariadb}
\label{fig:top-3c639f2-mariadb}
\vspace{-1ex}
\end{figure*}
\begin{figure*}[h]
\includegraphics[width=\textwidth]{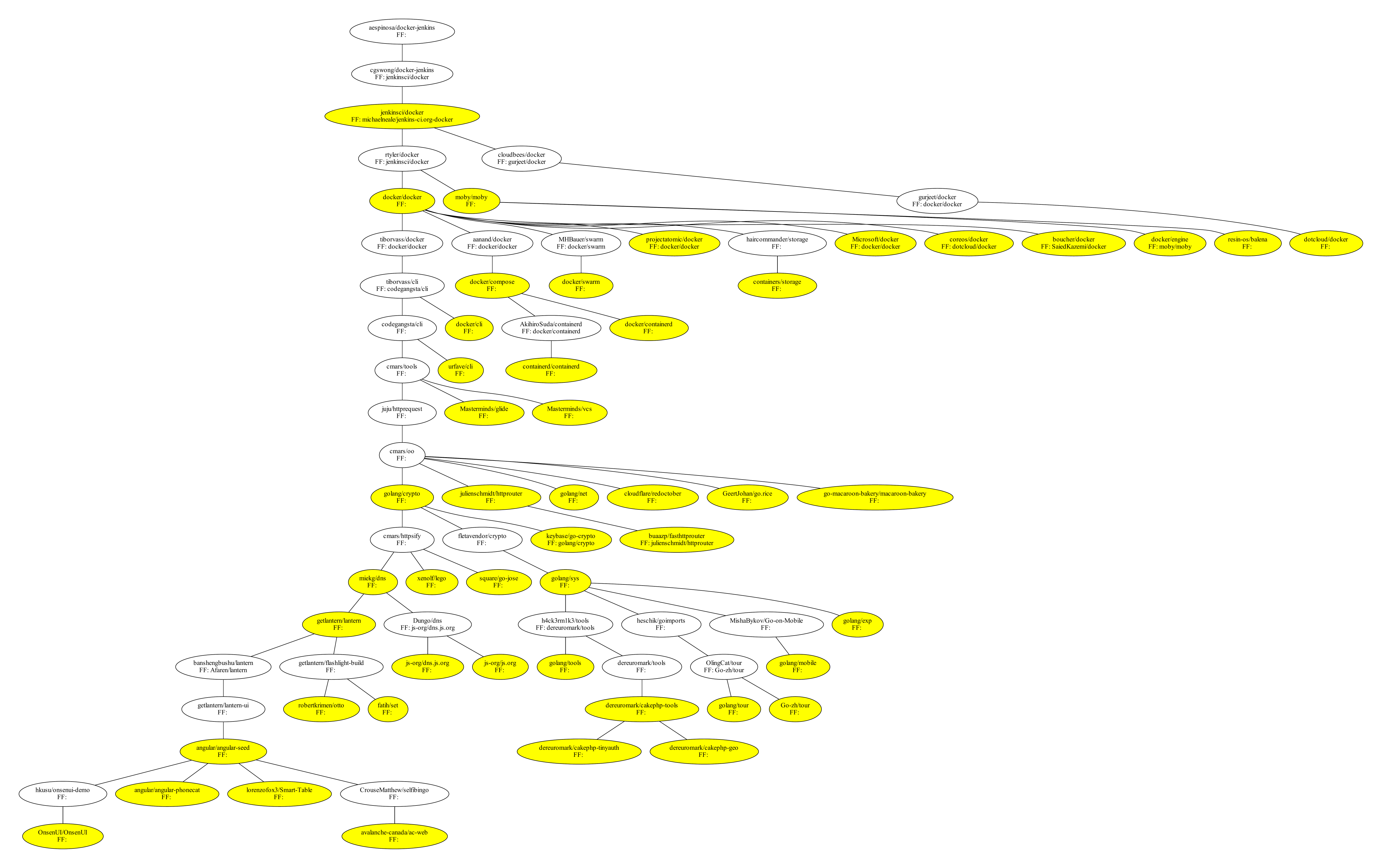}
\caption{Links associated with docker}
\label{fig:top-d516729-docker}
\vspace{-1ex}
\end{figure*}
\begin{figure*}[h]
\includegraphics[width=\textwidth]{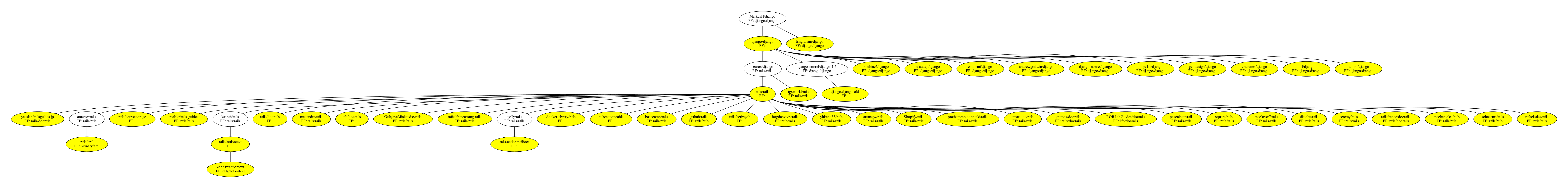}
\caption{Links associated with django-rails}
\label{fig:top-d516729-django-rails}
\vspace{-1ex}
\end{figure*}
\begin{figure*}[h]
\includegraphics[width=\textwidth]{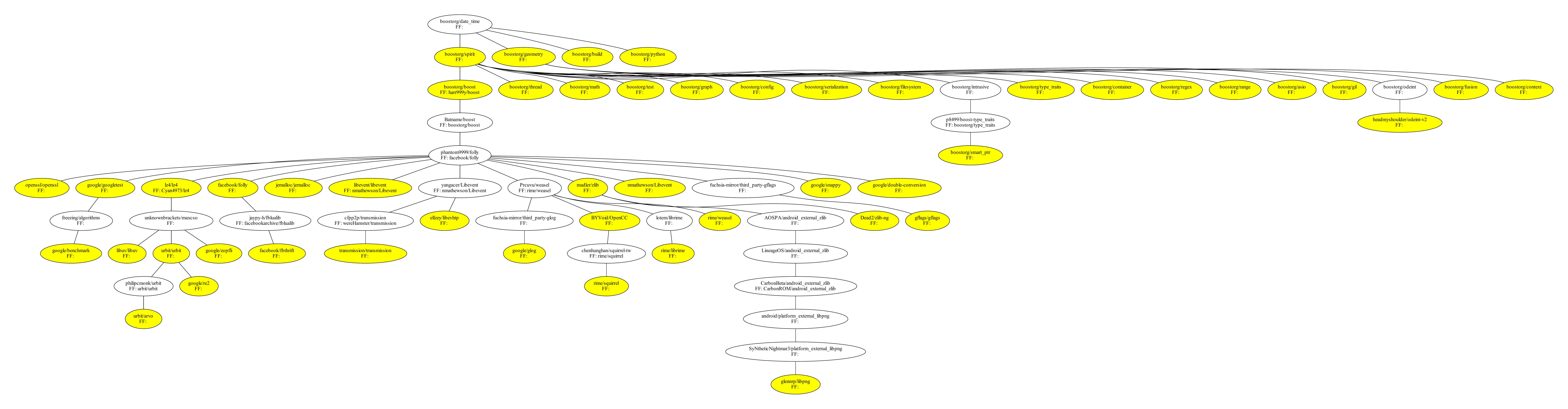}
\caption{Links associated with Google-zlib}
\label{fig:top-d516729-Google-zlib}
\vspace{-1ex}
\end{figure*}
\begin{figure*}[h]
\includegraphics[width=\textwidth]{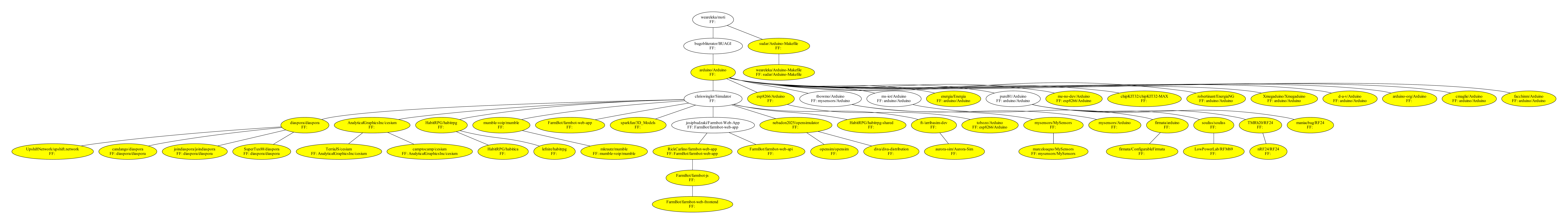}
\caption{Links associated with Diaspora-Arduino}
\label{fig:top-d516729-Diaspora-Arduino}
\vspace{-1ex}
\end{figure*}
\begin{figure*}[h]
\includegraphics[width=\textwidth]{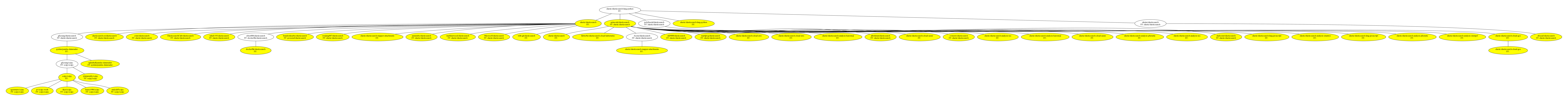}
\caption{Links associated with Elastic-Pandas}
\label{fig:top-d516729-Elastic-Pandas}
\vspace{-1ex}
\end{figure*}
\begin{figure*}[h]
\includegraphics[width=\textwidth]{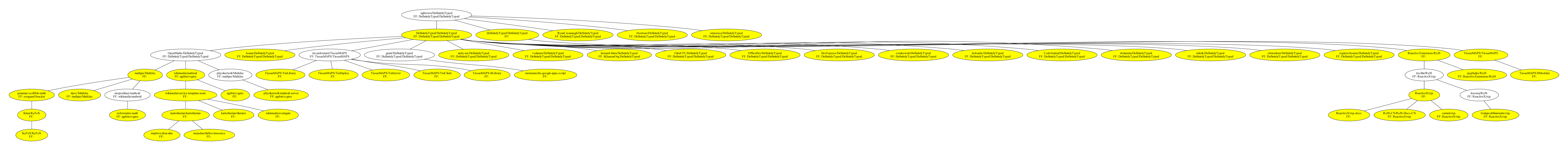}
\caption{Links associated with Definitely-Typed-RxJS}
\label{fig:top-d516729-Definitely-Typed-RxJS}
\vspace{-1ex}
\end{figure*}
\begin{figure*}[h]
\includegraphics[width=\textwidth]{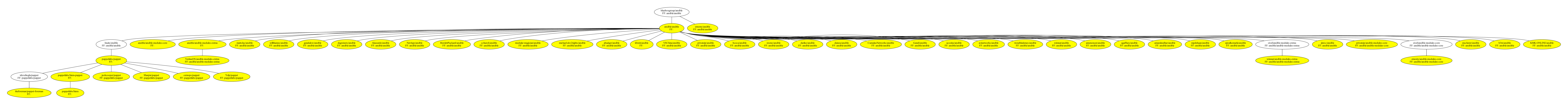}
\caption{Links associated with Ansible-Puppet}
\label{fig:top-d516729-Ansible-Puppet}
\vspace{-1ex}
\end{figure*}
\begin{figure*}[h]
\includegraphics[width=\textwidth]{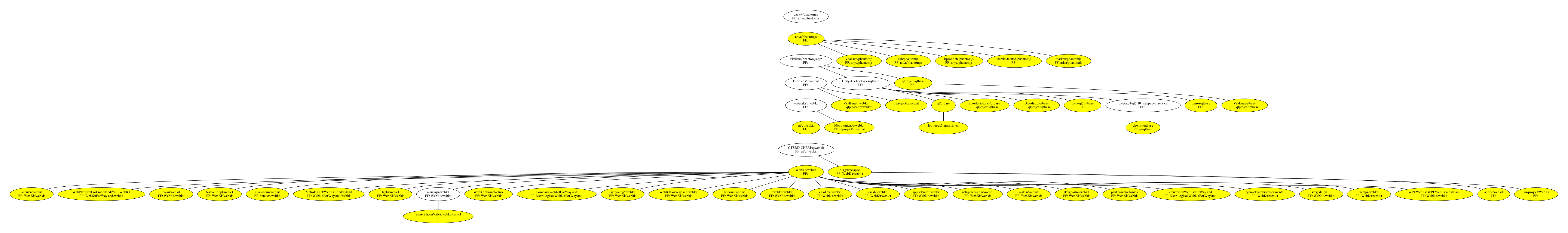}
\caption{Links associated with qt}
\label{fig:top-d516729-qt}
\vspace{-1ex}
\end{figure*}
\begin{figure*}[h]
\includegraphics[width=\textwidth]{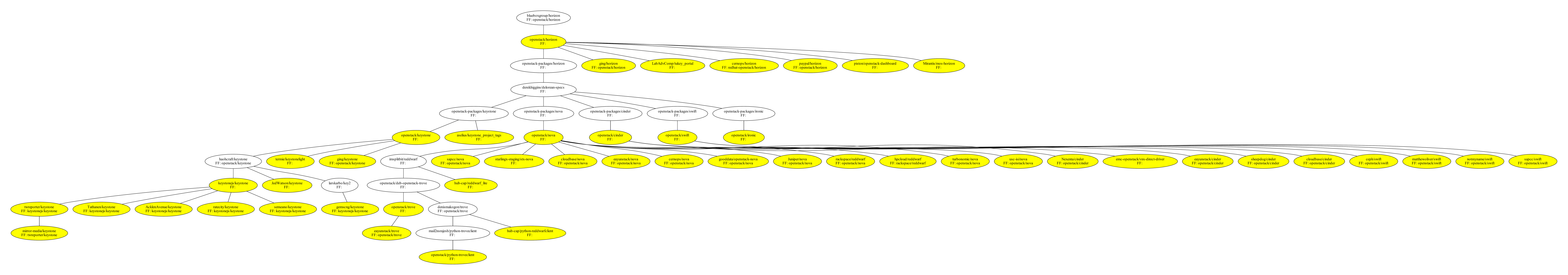}
\caption{Links associated with openstack}
\label{fig:top-d516729-openstack}
\vspace{-1ex}
\end{figure*}
\begin{figure*}[h]
\includegraphics[width=\textwidth]{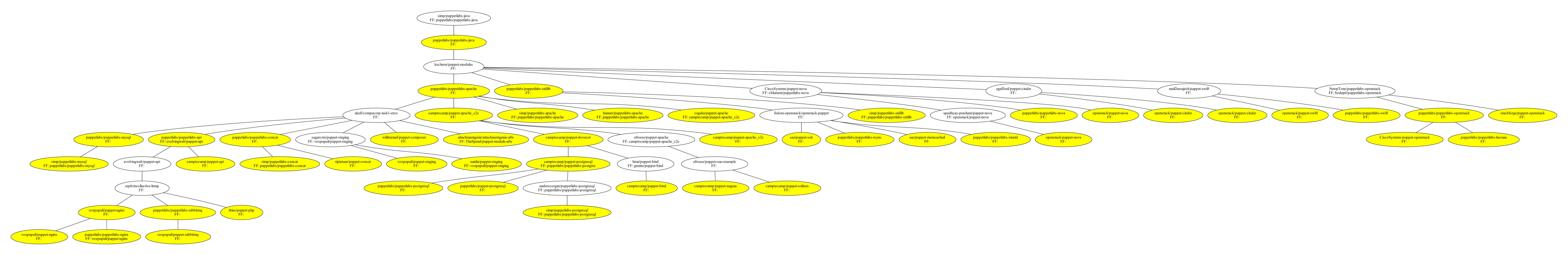}
\caption{Links associated with puppet-modules}
\label{fig:top-d516729-puppet-modules}
\vspace{-1ex}
\end{figure*}
\begin{figure*}[h]
\includegraphics[width=\textwidth]{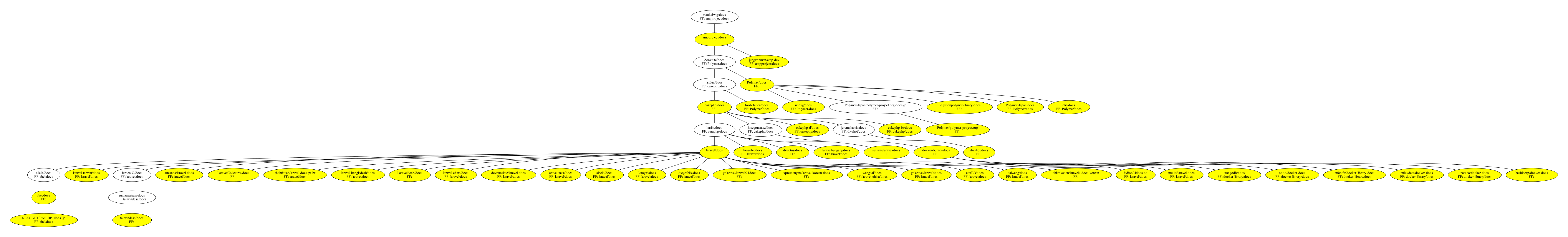}
\caption{Links associated with docs}
\label{fig:top-d516729-docs}
\vspace{-1ex}
\end{figure*}
\begin{figure*}[h]
\includegraphics[width=\textwidth]{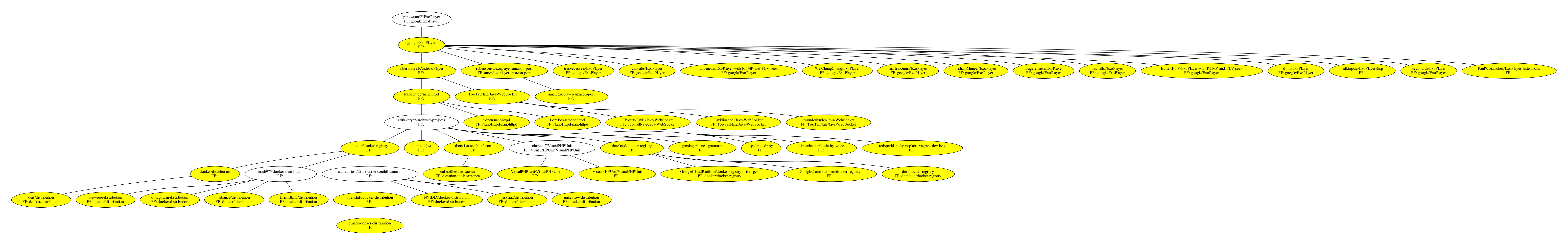}
\caption{Links associated with archived}
\label{fig:top-d516729-archived}
\vspace{-1ex}
\end{figure*}
\begin{figure*}[h]
\includegraphics[width=\textwidth]{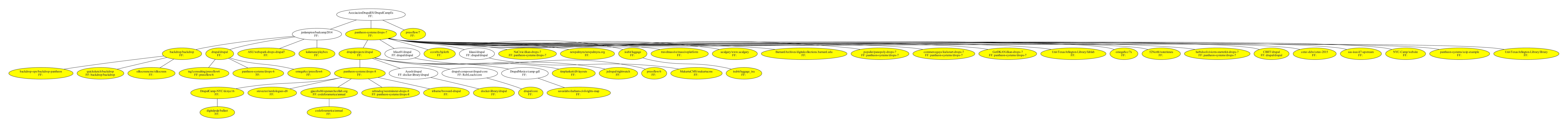}
\caption{Links associated with backgrop-drupal}
\label{fig:top-d516729-backgrop-drupal}
\vspace{-1ex}
\end{figure*}
\begin{figure*}[h]
\includegraphics[width=\textwidth]{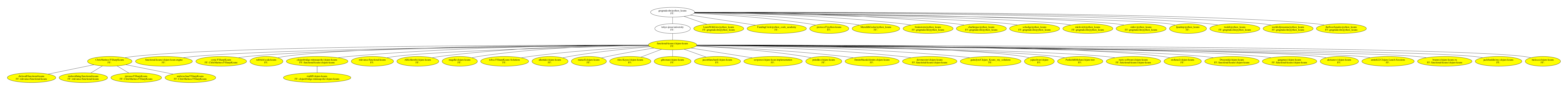}
\caption{Links associated with python-clojure-koans}
\label{fig:top-3c639f2-python-clojure-koans}
\vspace{-1ex}
\end{figure*}
\begin{figure*}[h]
\includegraphics[width=\textwidth]{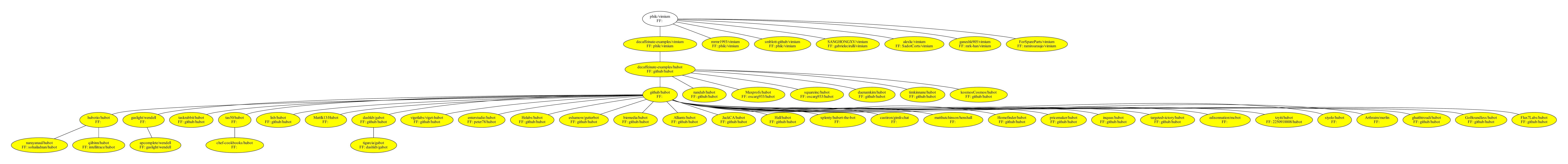}
\caption{Links associated with vimium-hubot}
\label{fig:top-3c639f2-vimium-hubot}
\vspace{-1ex}
\end{figure*}
\begin{figure*}[h]
\includegraphics[width=\textwidth]{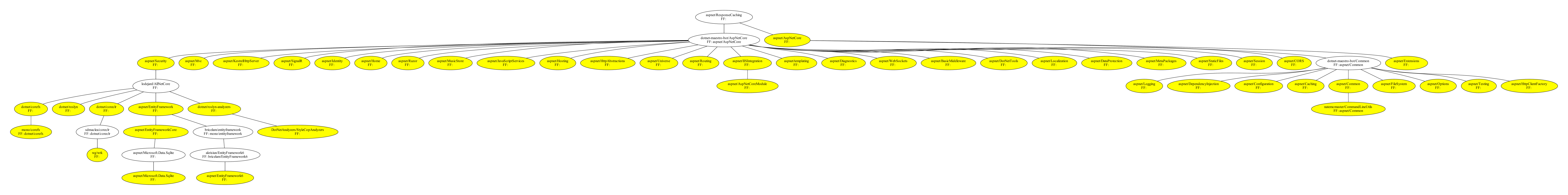}
\caption{Links associated with aspnet}
\label{fig:top-d516729-aspnet}
\vspace{-1ex}
\end{figure*}
\begin{figure*}[h]
\includegraphics[width=\textwidth]{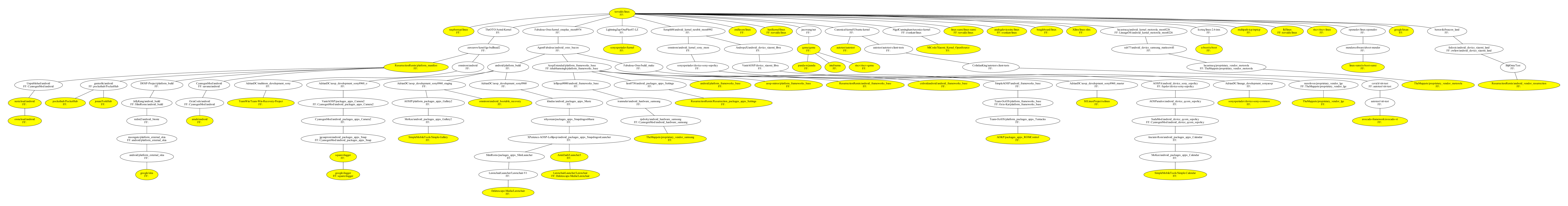}
\caption{Links associated with linux}
\label{fig:top-2b2b176-linux}
\vspace{-1ex}
\end{figure*}
\begin{figure*}[h]
\includegraphics[width=\textwidth]{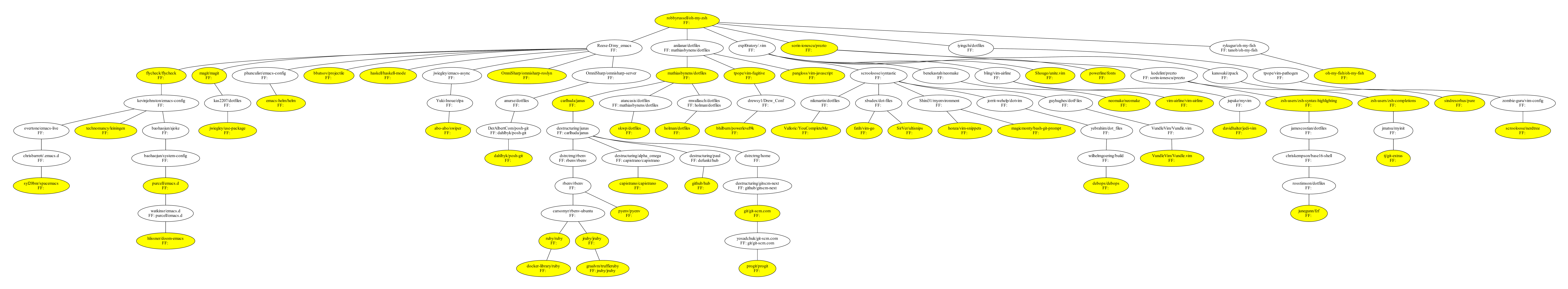}
\caption{Links associated with capistrano}
\label{fig:top-2b2b176-capistrano}
\vspace{-1ex}
\end{figure*}
\begin{figure*}[h]
\includegraphics[width=\textwidth]{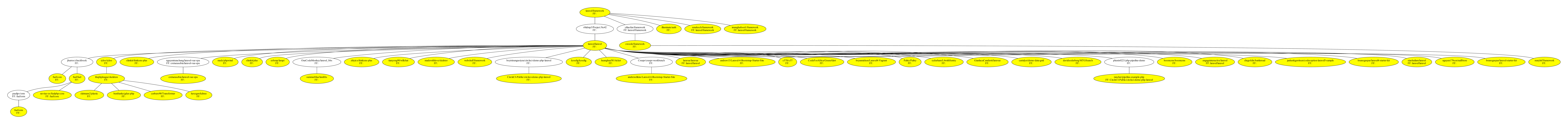}
\caption{Links associated with laravel-fuel}
\label{fig:top-2b2b176-laravel-fuel}
\vspace{-1ex}
\end{figure*}
\begin{figure*}[h]
\includegraphics[width=\textwidth]{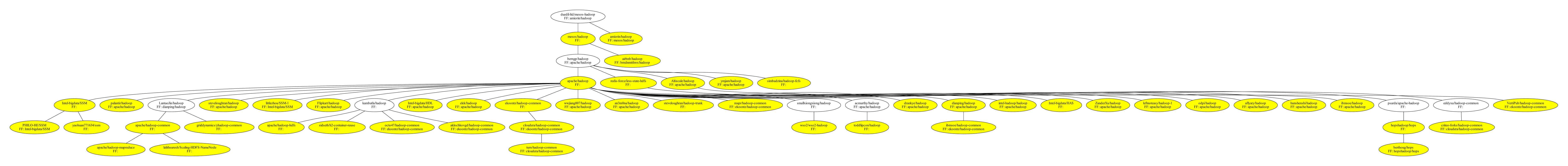}
\caption{Links associated with hadoop-ssm}
\label{fig:top-d516729-hadoop-ssm}
\vspace{-1ex}
\end{figure*}
\begin{figure*}[h]
\includegraphics[width=\textwidth]{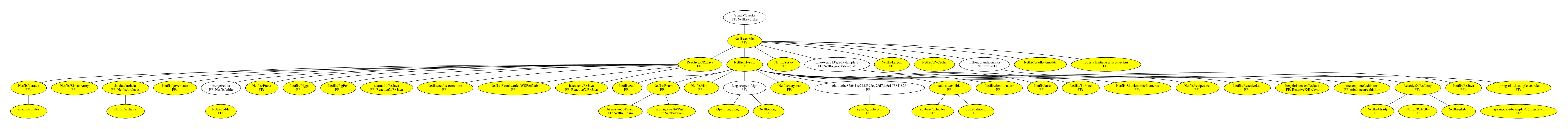}
\caption{Links associated with RcJava-Netflix}
\label{fig:top-d516729-RcJava-Netflix}
\vspace{-1ex}
\end{figure*}
\begin{figure*}[h]
\includegraphics[width=\textwidth]{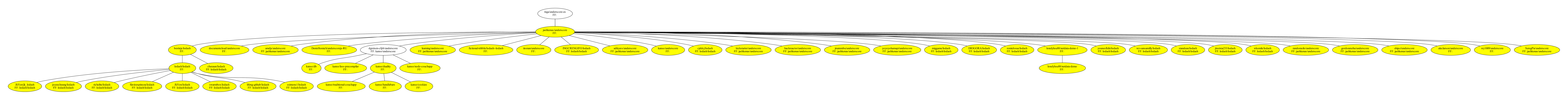}
\caption{Links associated with lodash-underscore}
\label{fig:top-d516729-lodash-underscore}
\vspace{-1ex}
\end{figure*}
\begin{figure*}[h]
\includegraphics[width=\textwidth]{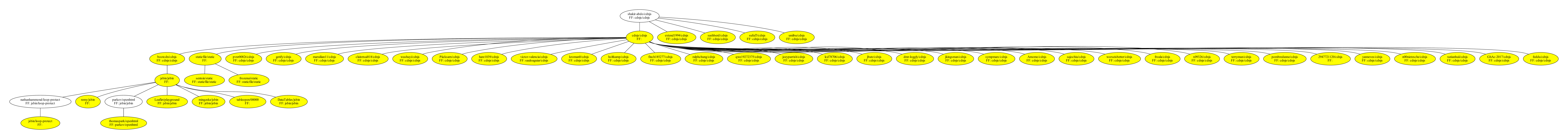}
\caption{Links associated with cdnjs-}
\label{fig:top-d516729-cdnjs-}
\vspace{-1ex}
\end{figure*}
\begin{figure*}[h]
\includegraphics[width=\textwidth]{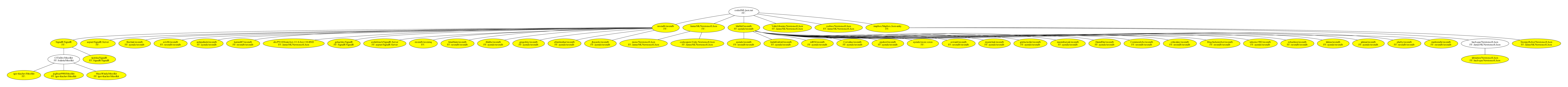}
\caption{Links associated with signalr-ravendb}
\label{fig:top-d516729-signalr-ravendb}
\vspace{-1ex}
\end{figure*}
\begin{figure*}[h]
\includegraphics[width=\textwidth]{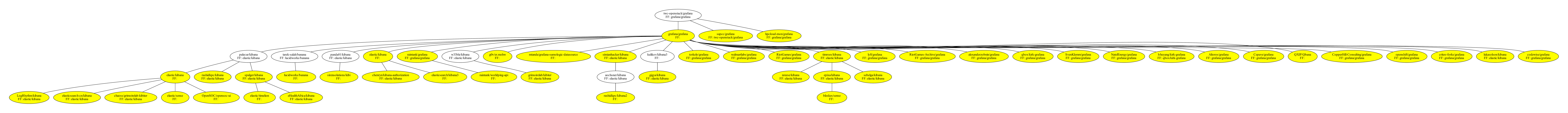}
\caption{Links associated with kibana-grafana}
\label{fig:top-d516729-kibana-grafana}
\vspace{-1ex}
\end{figure*}
\begin{figure*}[h]
\includegraphics[width=\textwidth]{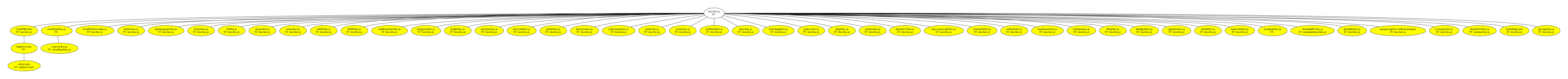}
\caption{Links associated with less-carto}
\label{fig:top-3c639f2-less-carto}
\vspace{-1ex}
\end{figure*}
\begin{figure*}[h]
\includegraphics[width=\textwidth]{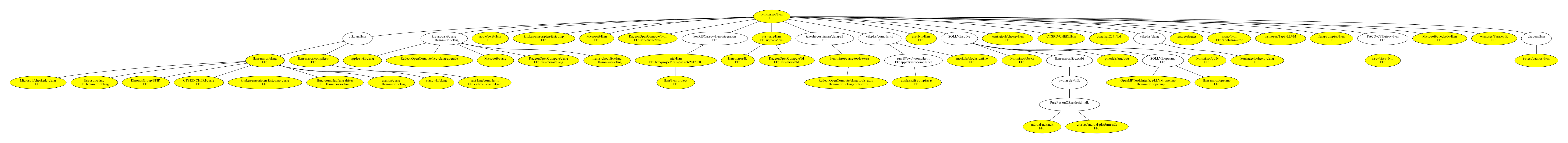}
\caption{Links associated with Swift-LLVM}
\label{fig:top-481f449-Swift-LLVM}
\vspace{-1ex}
\end{figure*}
\begin{figure*}[h]
\includegraphics[width=\textwidth]{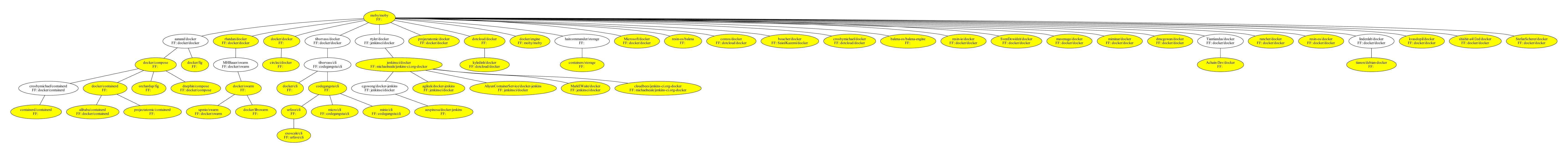}
\caption{Links associated with docker-containerd}
\label{fig:top-2b2b176-docker-containerd}
\vspace{-1ex}
\end{figure*}
\begin{figure*}[h]
{The figure's dimensions are too large for including it in a LaTeX document.
Open the figure in the paper's replication package.}
\caption{Links associated with disagreements}
\label{fig:disagreements}
\vspace{-1ex}
\end{figure*}

\fi

\end{document}